# Room Temperature Band-like Transport and Hall Effect in a High Mobility Ambipolar Polymer

*Satyaprasad P.Senanayak[1], A.Z.Ashar[1], Catherine Kanimozhi[2], Satish Patil[2*] and K.S.Narayan[1*]*

Affiliation:

[1] Chemistry and Physics of Materials Unit

Jawaharlal Nehru Centre for Advanced Scientific Research

Bangalore 560064, India

[2] Solid State and Structural Chemistry Unit,

Indian Institute of Science,

Bangalore 560012, India

*e-mail: narayan@jncasr.ac.in

satish@sscu.iisc.ernet.in

**Abstract:**

The advent of new-class of high-mobility semiconducting polymers opens up a window to address fundamental issues in electrical transport mechanism such as hopping between localized states versus extended state conduction. Here, we investigate the origin of ultra-low degree of disorder ($E_a \sim 16$ meV) and "band-like" negative temperature ($T$) coefficient of the field effect electron mobility: $\mu_{FET}^e(T)$ in a high performance ($\mu_{FET}^e > 2.5$ cm$^2$V$^{-1}$s$^{-1}$) diketopyrrolopyrrole (DPP)-based semiconducting polymer. Models based on the framework of mobility edge (ME) with exponential density of states are invoked to explain the trends in transport. The temperature window over which the system demonstrates delocalized transport was tuned by a systematic introduction of disorder at the transport interface. Additionally, the Hall mobility ($\mu_{Hall}^e$) extracted from Hall-voltage measurements in these devices was found to be comparable to field effect mobility ($\mu_{FET}^e$) in the high $T$ band-like regime. Comprehensive studies with different combinations of dielectrics and semiconductors demonstrate the effectiveness of rationale molecular design which emphasizes uniform-energetic landscape and low re-organization energy.

# 1. Introduction:

Organic electronics has received considerable attention due to its prospective potential in the embedded electronics space, driven by flexible low-cost solution processing and printing methods[1-2]. The performance and operating parameters such as field effect mobility ($\mu_{FET}$), switching speed and on-off ratio of these devices are ultimately governed by microscopic properties and the inherent chemical structure of the material[3]. It is well established that the degree of structural order and defect densities play a critical role in charge transport of organic semiconductors [4-7]. However, in contrast to inorganic semiconductors there is no commonly accepted model for charge transport in organic semiconductors. The general mechanisms of transport in case of semiconducting polymers has been largely dealt in the framework of hopping between disordered-localized states[8] and to a much lesser extent in terms of mobility edge (ME) models[9] where charges occupying density of states (DOS) above the mobility edge contribute to the conductivity. Polymer field effect transistors (PFETs) have attracted attention as a versatile platform to study charge transport in a controlled manner. PFET device architecture has been used to tune the conductivity ($\sigma$) of polymeric material from $10^{-5}$ S/cm to $10^{3}$ S/cm and observe different transport mechanisms [10-11] from localized activated behavior to band-like weak activated behavior[10,12-13]. From the classical theory of van der Waals solids[14], it is anticipated that band-like transport would be feasible if $\mu_{FET}$ exceeds 1 cm$^2$V$^{-1}$s$^{-1}$. Although, a range of polymers [15-16] have been demonstrated with $\mu_{FET}$ > 1 cm$^2$V$^{-1}$s$^{-1}$ which is comparable to organic crystals and a-Si, unambiguous evidence of band transport and ultra-low degree of disorder is not yet observed in these disordered systems. In addition, even the observation of apparent band-like behaviour in molecular systems is generally restricted to *p*- type materials[13]. We present results from an *n*-type DPP-polymer system which exhibits band-like transport and highlight the necessary prerequisites for this occurrence.

Diketopyrrolopyrrole (DPP) based molecules have attracted great attention due to high unipolar hole mobility ($\mu_{FET}^h$) ~ 10 cm$^2$V$^{-1}$s$^{-1}$ and balanced ambipolar mobility in a FET device[16]. The origin of high $\mu_{FET}$ in these polymers which lack long range order is an interesting open question[17]. In this study, we report a DPP based donor-acceptor system N-CS2DPP-OD-TEG (2DPP-TEG : **figure 1**) which shows high electron mobility ($\mu_{FET}^e$) of 2.5 cm$^2$V$^{-1}$s$^{-1}$ and balanced ambipolar transport. Beside the high $\mu_{FET}^e$, features such as ambient stability and reproducible characteristics unaffected by bias stress, makes this material a suitable candidate for fundamental studies regarding charge transport in polymers. In this study, we compare a range of dielectric and semiconducting materials to highlight the band-like negative *T* coefficient of $\mu_{FET}^e$ *(T)* and equivalence with $\mu_{Hall}^e(T)$ in the band regime of 2DPP-TEG based PFETs.

Solution processable films of small-molecules like TIPS-pentacene have exhibited trends of low activation magnitude and negative *T* coefficient of the *μ*$_{FET}$ at high gate voltages (*V$_g$*), which is attributed to localized transport limited by thermal lattice fluctuations rather than extended state conduction[7,18]. In solution deposited polymer thin-films, the microstructure is generally governed by the kinetics of film formation parameter and is expected to have higher degree of disorder. However, it is conceivable that semiconducting polymers based on appropriate molecular design and interfacial characteristics can provide a platform to investigate 2-D phenomena. In such systems, it may also be possible to suppress the transport fluctuations originating from structural disorder and traps by rationale molecular design[19]. The density of localized states can then approach the mobility edge and the low energetic disorder can be overcome by the prevailing thermal energy leading to a diffusive transport in continuum states. These characteristics can manifest as a negative temperature coefficient form of $\mu_{FET}(T)$ at high $T$. Interestingly, there have been no examples in polymeric systems which reveal features such as low activation energies ($E_a \sim 16$ meV) and crossover to an inverse dependence at high temperatures ($T > 200$ K).

In this report, we provide a clear evidence of one such ordered system obtained by an amphiphillic design of the molecule and demonstrate the effectiveness of the molecular design and interface engineering to introduce disorder in a controlled manner. The efficacy of molecular design is evident upon comparing the model system 2DPP-2Dod without the amphiphillic design which exhibits $E_a \sim 100$ meV and $\mu_{FET}^e \sim 10^{-3}$ cm$^2$V$^{-1}$s$^{-1}$, lower by three orders of magnitude than the amphiphillic counterpart. We also utilize the dielectric induced disorder and fluctuation at the interface to follow the $\mu_{FET}^e(T)$ behavior as a function of the broadening of DOS[20-22]. The transport mechanism gets altered from extended state transport to activated behaviour when the $k$ of the dielectric layer increases or when the underlying motif of amphiphillic molecular design is not present. In essence, 2DPP-TEG offers a plausible macromolecular model which has equivalence to an ordered organic small-molecule system.

## 2. Materials and Methods

**Materials**: Conjugated polymer 2DPP-TEG (weight average molecular weight, $M_w \sim 314,000$ g/mol) was synthesized as described elsewhere[15]. The dielectric materials hydroxyl free divinyltetramethylsiloxane bisbenzocyclobutene (BCB) was obtained from Dow Chemicals, PMMA ($M_w \sim 996,000$ g/mol) and PVDF-HFP copolymer pellets ($M_w \sim 455,000$ g/mol) were procured from Sigma Aldrich.

**Device Fabrication**:

**a) PFET** : Bottom gated top contact PFETs were fabricated by coating Al/Cr-Au ($10^{-6}$ mbar, 1 A$^0$/s, 30 nm thick) by shadow mask technique on standard RCA cleaned glass substrates. A dielectric layer was obtained by spin coating the solutions from their respective solvents (PVDF-HFP at 80mg/ml in N,N-Dimethylacetamide, PMMA in propylene carbonate at 80 mg/ml and BCB in Mesitylene) at 1000 rpm for 1 minute to obtain films of thickness 0.2 - 0.4 µm. The dielectric films were annealed in $N_2$ atmosphere. 2DPP-TEG layer is spin coated from a solution of 10mg/ml in chlorobenzene at 1000 rpm for 1 minute to obtain films of thickness ~ 80 nm. Optimized annealing condition to obtain the best performance for this polymer was obtained as $180^0$ C for 2 hrs. This was followed by the deposition of Au or Al source drain electrode ($10^{-6}$ mbar, 1 A$^0$/s, 20 nm thick) by shadow masking technique to obtain channels of length 5 - 100 µm and width 1 mm for ambipolar or n-channel PFETs respectively. For vacuum gated devices a spacer layer of ~ 200 nm is introduced around the gate electrode and polymer film with S-D electrodes is placed on it to complete the device. Similar procedure was also used to fabricate devices with 2DPP-2Dod molecule.

**b) Diode for SCLC**: Devices were fabricated on pre-cleaned ITO substrates (15 ohm/□), obtained from XINYAN Technology Ltd. For hole only devices (ITO/PEDOT:PSS/Polymer/Au), PEDOT:PSS was spin coated at 2000 rpm and annealed at $110^0$C for 1 hour in air. Polymer films of varying thickness (200 nm to 2 µm) were obtained and annealed at $180^0$C for 2 hours in $N_2$ atmosphere. Similarly electron only devices were fabricated on Al coated pre-cleaned glass substrates on which polymer layer is coated followed by Al cathode electrode. Metal electrodes were coated by thermal evaporation ($10^{-6}$ mbar, 1 A$^0$/s, 60 nm thick).

**c) Inverter fabrication**: Complementary inverters were fabricated as top-contact bottom gate structure with 2DPP-TEG layer acting as both p-channel and n-channel. The metallic interconnects between the transistors were coated by thermal evaporation of Au after the fabrication of individual transistors. Noise margin and gain was optimized by varying $W_L/W_D$ ratio.

**Electrical characterization**: PFET characterization was carried out using two source measuring units - Keithley 2400 and high impedance electrometer Keithley 6514. Capacitance measurements were done at 100 Hz using HP4294A. SCLC measurements were performed using Keithley 2400 and high impedance electrometer Keithley 6514 and the T variation was carried out using a closed cycle Helium Cryo set-up (CTI inc.).

**Hall measurement**: Bottom gated top contact PFETs were fabricated using transparent patterned ITO (15 ohm/□, obtained from XINYAN Technology Ltd.) as the gate. Transparent gate electrodes were chosen to ease the patterning of the subsequent channel and Hall electrodes. BCB dielectric layer was then spin coated at 1000 rpm for 1 minute and annealed in $N_2$ atmosphere to obtain films of thickness 0.4 μm. Patterned semiconducting layer is introduced using lithographic and printed techniques to accommodate only the S-D electrodes and the Hall probes. The patterning of the semiconducting layer minimizes leakage pathways being picked up by the Hall probes. This was followed by the deposition of Al source drain electrode ($10^{-6}$ mbar, 1 $A^0$/s, 20 nm thick) by shadow masking technique to obtain channels of length 200 - 400 μm and width 200 – 500 μm. Al electrodes transverse to the channel were coated ($10^{-6}$ mbar, 1 $A^0$/s, 20 nm thick) to serve as the Hall probes using physical masking technique after aligning them under the microscope. The fabricated devices had dimensions in the range of L ~ 200 - 400 μm, W ~ 200 – 500 μm, L* ~ 200 μm – 1 mm and W* ~ 40 – 80 μm.

Hall measurements were performed using a home-built probe assembly in a physical property measurement system from Quantum Design Inc where **B** is swept in the range of 8 T to - 8 T in steps of 0.1 T. Keithley 2001 Multimeter (input impedance ~ 10 TΩ) was used to measure $V_H$ while the PFETs were biased ($V_d$ = 10 V and $V_g$ = 60 V) using Keithley 2400 source meters. To calibrate and verify Hall measurements additional standard in-organic samples with known Hall voltage values were used in the same experimental setup.

**AFM measurement**: For the imaging, JPK Nanowizard 3 with APPNANO ACTA Aluminium coated Si-cantilevers (force constant 40 N/m and resonance frequency f ~ 375 kHz) were used in non-contact mode.

**DFT calculation**: To optimize the molecular structure, DFT calculations were performed on 2DPP-TEG oligomers (n = 5 repeat units) using the hybrid exchange correlation functional (B3LYP) 6-31G** basis set with Gaussian-09 package. The molecular structure is optimized both in neutral and in the charged (anionic) state to obtain an estimate of electron re-organisation energy.

## 3. Results and Discussion

### 3.1. Temperature dependent $\mu_{FET}^e$ measurement

DPP based bottom gate (G) top contact (S-D) FETs were fabricated on a range of dielectric materials (varying dielectric constant-$k$) to study the transport behavior. Transistors with distinct linear and saturation regime in $I_{ds}(V_d)$ profiles and minimal hysteresis in $I_{ds}(V_g)$ characteristics form appropriate representative of the devices (**Figure: 1a**). Devices with Al S-D electrodes and BCB dielectric layer demonstrated reliable leakage free behaviour and exhibited $\mu_{FET}^e$ ~ 2.5 cm$^2$V$^{-1}$s$^{-1}$ in the saturation regime as estimated from the standard transconductance equation. Balanced ambipolar charge transport with maximum $\mu_{FET}^e$ ~ 0.46 cm$^2$V$^{-1}$s$^{-1}$ and $\mu_{FET}^h$ ~ 0.41 cm$^2$V$^{-1}$s$^{-1}$ were obtained with Au S-D electrodes in the saturation regime. The ON-OFF ratios exhibited by these PFETs were typically in the range of 10$^5$ - 10$^6$. Measurement of $\mu_{FET}^e$ and $\sigma$ were performed over a $T$ range of 40 K – 300 K on a variety of dielectrics as shown in **Figure: 2a**. The dielectrics covered the category of low-$k$, high-$k$ and relaxor materials. Considering the long channel length ($L \approx 60$ μm - 100 μm) used in these devices, contact resistance is negligible compared to channel resistance and does not affect the transport measurements significantly. The key findings of the $T$ dependent measurements are: (i) ultra-low level of energetic disorder signified by the magnitude of activation barrier; (ii) crossover from $\frac{d\mu^e_{FET}}{dT} > 0$ at low $T$ (40 K – 200 K) to $\frac{d\mu^e_{FET}}{dT} < 0$ at high $T$ (> 200 K); (iii) tunability of the transport mechanism from activated to adiabatic by modulating the interface disorder, charge density and molecular engineering.

It is instructive to examine the $\mu^e_{FET}(T)$ behavior using simple single activated fits to gauge the magnitude of thermal barrier. Transport measurements indicate low $E_a \approx 16$ meV (in $T$ range 100 K < $T$ < 220 K) for PFETs fabricated with BCB dielectric and Al S-D electrodes. As the dielectric strength $k$ increases from 2.6 to 8 (for PVDF-HFP), the maximum $\mu^e_{FET}$ (in activated regime) decreases from 2.5 cm$^2$V$^{-1}$s$^{-1}$ to 1.5 cm$^2$V$^{-1}$s$^{-1}$ and $E_a$ increases to 47 meV. A characteristic of significance in the $\mu^e_{FET}(T)$ profiles is the transition from $\frac{d\mu^e_{FET}}{dT} > 0$ at low $T$ to $\frac{d\mu^e_{FET}}{dT} < 0$ at high $T$, around a crossover temperature $T_{trans}$. $\mu^e_{FET}(T)$ measurements indicate $T_{trans}$ ~ 200 K for PFETs fabricated with BCB dielectric layer which increases to around 220 K for PFETs with PMMA as the dielectric layer. However, PFETs fabricated with relaxor dielectric layer PVDF-HFP do not exhibit any crossover in the entire $T$ range. This behavior can be understood based on a simple argument that in order to observe a crossover to adiabatic transport in molecules the energy bandwidth should be larger than the energy change involved in scattering, which implies the minimum condition of mobility $(\mu_{min}) > \frac{er^2}{2\hbar}$, where $r$ is the intermolecular distance, $e$ is the electronic charge and $\hbar$ is the reduced Plank's constant[14]. In case of PVDF-HFP based PFETs, $\mu^e_{FET}$ is less than the required $\mu_{min} \approx 1.9$ cm$^2$V$^{-1}$s$^{-1}$ (for $r$ = 0.36 nm for 2DPP-TEG obtained from GIXRD). Absence of a transport cross-over in PVDF-HFP based PFETs can then be related to the lower $\mu_{FET}^e$ magnitude originating from dipolar induced disorder as depicted in Schematic: 1b.

Reported observations of band-insulator transition in polymers/organic molecule have been attributed to factors like: (i) field induced crossover in conductivity[11,23] or (ii) Luttinger liquid type 1-D metallic conductivity[24]. In the present case, the FET output characteristics demonstrated well defined linear and saturation regime without any non-linear transport behavior even at $T \approx 40$ K. This indicates that the transport is relatively independent of the lateral field unlike other high mobility polymers where lateral field causes de-trapping of carriers from shallow traps. Hence, the possibility of field induced cross-over from activated to field emission can be neglected in the present case. In addition, we observe a clear Hall effect (as described in later section) as well as interconnected network of self-assembled polymeric structures (**Figure: 3**) which supports a 2-D microstructure for transport and does not indicate a Luttinger-liquid type mechanism in this system. A phenomenological justification of 2-D transport based on variable range hopping (VRH) in the localized state for $T < T_{trans}$ is provided in **Supplementary Section S1**. It has also been reported that negative coefficient of $\mu_{FET}(T)$ in tetracene like molecular systems arises from a $T$ induced structural re-arrangement[25] or due to bias stress in the molecule[26]. Factors relating to any structural transition are discounted in the present case based on DSC and FTIR measurements which do not indicate any structural transition or abnormal water related peaks in the entire $T$ range (details in **Supplementary Section S2**). Similarly factors like bias stress as the origin of the $\mu_{FET}^e(T)$ trends were also discounted, since the $\mu_{FET}^e(T)$ trend is retained in the forward and reverse temperature cycle. Moreover, the transconductance curve obtained at $V_d = 60$ V is stable over a duration greater than $10^4$ s (**Supplementary Section S3**). In addition, the non-linear behavior in $\mu_{FET}^e(T)$ has also been attributed to water absorption[6,27-28]. We provide evidences in **Supplementary Section S4** that clearly does not support the possibility of water related traps as a possible origin for $\mu_{FET}^e(T)$ trends. The reproducible features obtained from measurements of large number of devices and different batches of samples strongly suggest the inherent electronic origin representing the characteristics of the system.

### 3.2 Analysis of the disorder in charge transport

The ME framework offers a scenario where the cross-over in transport is a continuous transition, unlike the hopping based models where crossover is accompanied by a discontinuity in transport parameters[29]. Evidence such as $E_a \approx k_B T_{trans}$ offer a simple argument supporting ME based processes. In order to obtain the DOS profiles, $\mu_{FET}^e(V_g)$ was measured and analyzed. It was observed that $\mu_{FET}^e$ increases rapidly at low $V_g$ (> 40 V) as the states are quickly filled up and at high $V_g$ (≥ 40 V), $\mu_{FET}^e$ becomes independent of the induced charges at the interface corroborating the existence of exponential-type DOS. The magnitude of $E_a$ and the nature of the DOS indicates that the trap levels are close to the band edge which explains the origin of delocalized transport due to de-trapping of charge carriers by thermal fluctuations[19]. The effective mobility $\mu_{FET}^e$ is expressed as $\mu_0 \frac{\tau}{\tau + \tau_{tr}}$, where $\tau$ is the average time the carrier spends travelling between shallow traps, $\tau_{tr}$ is the lifetime of shallow traps and $\mu_o$ is the trap free mobility. If $\tau \ll \tau_{tr}$ the transport is trap limited with activation energy as obtained from Arrhenius fit, however in the regime $\tau \gg \tau_{tr}$ corresponding to the extended transport regime close to the mobility edge $\mu_{FET}^e \sim \mu_0 T^\gamma$ with γ ~ 0.9 - 1.4 (Table 1) [30]. The magnitude of γ is reflective of the prevailing scattering contributions from impurities, which in the present molecule is relatively lower than other organic molecules like rubrene and pentacene[31-32].

The observed dependence of $\mu_{FET}^e$ on different dielectric layers (**figure: 2a**) can also originate from the variation of the interfacial width[33]. To discount this possible interpretation, high resolution AFM was utilized to measure the roughness of the dielectric layers before and after the solvent treatment. AFM profiles shown in **Supplementary Figure: S6** do not exhibit significant increase in surface roughness of the dielectric layer with solvent treatment, indicating minimal effect of interface de-mixing on the field effect transport. On this basis, the $\mu_{FET}^e(k)$ behavior can be attributed to the dipolar disorder of the dielectric layer. The magnitude of polar disorder induced due to the dielectric ($E_p$) is estimated from the expression: $E_a \sim \sqrt{E_{sc}^2 + E_p^2}$, where $E_{sc}$ is the activation in the low-$k$ BCB dielectric layer based PFET which is assumed to have negligible $E_p$[34]. The dipolar induced broadening in the DOS of 2DPP-TEG originating from the dielectric layers PMMA and PVDF-HFP is $\approx$ 26.5 meV and 44.3 meV respectively (details in **Supplementary Section S6**) which is considerably lower than the estimates obtained for amorphous polymers like PTAA[20]. This relatively lower dipolar-broadening contribution can be related to electrostatic isolation of the conjugated core due to edge-on stacking of the polymer on the dielectric surface as seen in the GIXRD (**Supplementary Section S7**). Nevertheless, this low but finite dielectric disorder appears to be critical in eventually deciding the $\mu_{FET}^e(T)$ behaviour.

### 3.3. Effect of interface charge density on the transport mechanism

To understand the role of gate induced carrier density ($n_G$) on the order-disorder transition, $\mu_{FET}^e(T)$ is measured at different $V_g$ (**Figure: 2b**). PFETs fabricated with BCB dielectric layer and 2DPP-TEG molecule do not demonstrate a transport crossover for low $V_g$ ($\leq$ 20 V) however, at higher $V_g$ ($\geq$ 40 V), $T_{trans}$ gets prominent indicating the crossover. Similarly, for $V_g \leq$ 20 V, $\mu_{FET}^e$ is dependent on $V_g$ and for $V_g \geq$ 40 V, the $\mu_{FET}^e$ (at 300 K) varies in the range of 1.4 $\pm$ 0.1 cm$^2$V$^{-1}$s$^{-1}$ for a $V_g$ change from 40 V to 80 V indicating a weaker $V_g$ dependence of $\mu_{FET}^e$. This corroborates with the weak $E_a$ and low energetic disorder for charge transport in the PFETs fabricated with BCB dielectrics. In the case of PFETs fabricated with PVDF-HFP dielectric layer no crossover is observed in the $\mu_{FET}^{e}(T)$ trends even at $n_G \sim 4 \times 10^{19}$ cm$^{-3}$ which can be attributed to high degree of interfacial disorder ($E_a \sim 47$ meV). The importance of the dielectric-semiconductor interface and the annealing effects becomes apparent upon comparison with vacuum-gated devices. Vacuum-gated devices show relatively low $\mu_{FET}^{e}$ of $10^{-3}$cm$^2$V$^{-1}$s$^{-1}$ and higher $E_a \sim 150$ meV indicating the presence of large number of electron-traps at the interface. These performance parameters are comparable to other vacuum gated polymeric devices[35] and can be related to the presence of un-passivated surface trap states in these structures (details in **Supplementary Section S8**) due to the significantly low $n_G$. These observation can be elucidated in the framework of ME model for charge transport. Charge transport in polymeric materials involves thermal excitation to states at or below the transport level ($E_t$). As the states above $E_t$ are filled up with increase $n_G$ the quasi-Fermi level is lowered and $E_a$ decreases[35]. Thus extended transport in polymeric materials requires a combination of an ordered interface as well as sizeable amount of $n_G$ to passivate the traps.

**3.4. Atomistic origin of disorder in 2DPP-TEG**

At a microscopic level, the origin of high $\mu_{FET}^{e}$ and low degree of disorder in 2DPP-TEG can be justified from computational studies at the molecular scale. Quantum chemical calculations were carried out using density functional theory (DFT) with the B3LYP hybrid functional and 6-31G** basis set. The energy optimized structure demonstrated electron transfer Marcus re-organization energy ($\lambda_e$) of 0.11 eV and a torsional disorder of ≈ 2° from a perfectly co-planar conjugated core. Despite the limited number of oligomers (5) considered and neglecting the polarization effect introduced by the dielectric media, $\lambda_e$ values are qualitatively consistent with the experimentally observed $E_a$. From the semi-classical Marcus theory the activation energy for the self-exchange charge-transfer reaction is given by $\lambda_e/4$[36]. The contribution of $E_a$ is generally overwhelmed by the inherent disorder of the polymeric systems. Interestingly, in this system $\lambda_e/4$ (25.5) $\approx E_a$ (16 meV) indicating that the activation energy required for de-trapping of charge carriers originate from the activation mechanism involved in conformational changes. This observation is consistent with observed high mobility values and low degree of disorder for charge transport.

In addition, 2DPP-TEG thin films demonstrates high degree of crystallinity and enhanced $\pi$-$\pi$ stacking due to stronger aggregation which originates from amphiphillic design of molecule[5,17]. Analysis of the morphology by high resolution AFM (details in experimental section) showed an interconnected network of self-assembled structures (**Figure: 3a**) making the transport more tolerant to disorder. This is consistent with the observation of low $E_a$ (~ 16 meV) in the temperature dependent transport. Furthermore, the observed self-assembled structures are retained for films of varied thickness pointing to the presence of the interconnected self-assembled structures throughout the bulk of the films which significantly contributes to the observed disorder free transport mechanism. In order to analyze the role of $\pi$-$\pi$ stacking and the triethylene glycol (TEG) substitution, PFETs were fabricated under similar condition with a control polymer 2DPP-2Dod. In the case of 2DPP-2Dod where the amphiphillic interactions are absent, the interconnected self-assembled structures were not present (**Supplementary Section S9**). It was noted that this analogue of DPP without the order-promoting TEG group demonstrated $\mu_{FET}^e$ ~ $10^{-3}$ cm$^2$V$^{-1}$s$^{-1}$ and $E_a \approx$ 100 meV in the regime < 280 K without any crossover in the $\mu_{FET}^e(T)$ trends. These transport measurements clearly points to the role of the amphiphillic design in promoting morphological and energetic order for efficient charge transport. Hence, presence of strong aggregation due to efficient $\pi$-$\pi$ stacking, interconnected network of ordered regions which provide efficient pathway for transport within the disordered regions and co-planarity of the conjugated core are possibly the reasons for 2DPP-TEG to be more tolerant to disorder.

### 3.5. Hall measurement

In the band-like transport regime ($T > T_{trans}$) to confirm the presence of band-like conduction, Hall voltage ($V_H$) measurements were performed. $V_H$ signatures in a PFET is a sensitive probe to measure the degree of delocalized adiabatic transport, since it requires the charges to be mesoscopically extended to have a well-defined wave vector which can couple to the magnetic field[30-31]. Observation of clear $V_H$ in PFETs have been challenging due to low $\mu_{FET}$, high contact resistance and threshold voltage instabilities, moreover obtaining clear negative $T$ dependence is relatively rare in disordered materials. For the $V_H$ measurement, Hall probes were coated on FETs fabricated with patterned semiconductor layer (as shown in top inset **Figure: 4a**). Typical device dimensions were maintained in the range of $L$ = 200 - 400 μm, $W$ = 200 - 500 μm and the distance between Hall electrodes were varied in the range of 200 μm - 1 mm with width of 40 μm - 80 μm (device fabrication in experimental section). This device geometry is similar to the Hall probe assembly used for a-Si and rubrene[37-39]. $V_H$ signal was observed in devices at 300 K with BCB dielectric layer and Al S-D electrodes for $|V_g|$ > 60 V and $V_d$ = 10 V. The observation of $V_H$ signal is consistent with the $\mu_{FET}^e(V_g)$ trend, wherein at high $V_g$ the transport is trap-free and independent of induced charge density. Experimental artefacts were minimized by measuring $V_H$ as a function of ***B*** over a wide range (0 T to 8 T) as shown in **Figure: 4a**, and subsequently verifying the trend by reversing the magnetic field direction and changing the polarity of $I_{ds}$. Typical $V_H$ signal over multiple cycles of ***B*** is provided in the **Supplementary Section S10**. The trap free charge density extracted from the slope of $V_H$ variation with ***B*** ($n_H = I_{ds}B/eV_H \approx 1.4 \times 10^{12}$ cm$^{-2}$) is in agreement with the surface charge density estimated from gate capacitance measurements ($n_{FET} = C_i (V_g - V_{th})/e \approx 1.2 \times 10^{12}$ cm$^{-2}$ - $1.6 \times 10^{12}$ cm$^{-2}$) indicating a conclusive evidence for extended state transport. Remarkably, the magnitude of $\mu_{Hall}^e$ was comparable to $\mu_{FET}^e$ in the delocalized transport regime ($T > T_{trans}$) and falls steeply in the trap-limited activated transport regime ($T < T_{trans}$). The magnitude of $V_H$ increases with $V_g$ which is consistent with the observed trends of $\mu_{FET}^e(V_g)$. Ratio of $n_H/n_{FET}$ can be used as a measure of the fraction of delocalized charge carriers; this ratio increases with increase in $T$ (in band-like transport regime) and assumes a value close to unity at 300 K. A smaller magnitude of $V_H$ with $n_H << n_{FET}$ is observed at low $T$ ($< T_{trans}$) of 200 K corresponding to the low $T$ hopping regime where quantum mechanical interference among different hopping processes can contribute[37]. The key aspect is the equivalence of intrinsic trap-free $\mu_{Hall}^e$ and $\mu_{FET}^e$ at high $T$ in 2DPP-TEG system bringing out the general consistency of transport occurring in delocalized states

### 3.6. Bulk transport measurement

Detailed picture of the bulk disorder is essential to completely understand the origin of extended state conduction in 2DPP-TEG. Further studies were carried out to determine the bulk properties from space-charge limited (SCL) measurements. For electron only devices Al/polymer/Al device architecture was used whereas for hole only devices ITO was used as anode and Au was used as the cathode. Current density as high as $10^3$ A/m$^2$ is typically obtained for films of thickness 0.5 μm at a low bias ~ 10 V. Clear and distinctive linear and SCL regimes were obtained for both *p*-type and *n*-type transport as seen in the **Figure: 5**. The bulk mobility ($\mu_{SCL}$) was obtained using the Mott-Gurney equation given by: $J_{SCL} = \frac{9}{8}\varepsilon_0 k_S \mu_{SCL} \frac{V^2}{d^3}$, where $J_{SCL}$ is the current density of the device in the SCL regime, $k_s$ ($\approx$ 3 - 4) is the relative permittivity of the semiconductor, *d* is the thickness of the active semiconducting layer and *V* is the bias applied. Typical $\mu_{SCL}$ magnitude of $(9 \pm 5) \times 10^{-3}$ cm$^2$V$^{-1}$s$^{-1}$ was obtained with electron only devices and $\mu_{SCL}$ of $(6 \pm 2) \times 10^{-3}$ cm$^2$V$^{-1}$s$^{-1}$ was obtained with hole only devices. Activation energy involved in the bulk transport was obtained from the *T* dependent transport measurements performed in the *T* range of 100 K – 300 K. The presence of a clear SCL regime at low *T* points to the absence of any deep-trap dominated non-linear processes in the bulk transport. Activation energy from the SCL ($E_A^{SCL}$) measurements were 60 ± 5 meV for *n*-type transport which is comparable to reported high mobility *n*-type polymers[40]. However, *p*-type devices indicate a stronger activation in $\mu_{SCL}(T)$ with $E_A^{SCL}$ ~ 135 ± 12 meV. Devices were also fabricated with polymer layer thickness varying from 0.2 μm - 2 μm to understand the limiting factors in the bulk transport. In the SCL regime it is expected that current density scales as *J(E)* ∝ *d$^{-1}$* whereas, if the current is injection limited(IL) *J(E)* is independent of the semiconductor thickness[40]. In the case of *n*-type diodes, *J(E)* has a linear variation with 1/*d* indicating SCL transport. Interestingly, for *p*-type diodes *J(E)* is independent of film thickness in the range of 0.2 μm to 1 μm and has a linear dependence for films of thickness > 1 μm, indicating a crossover from IL behavior to SCL behavior with increase in active layer thickness. Bulk transport measurements demonstrate the underlying difference in the disorder energetics of the hole and electron transport in this polymer which is consistent with the trends in field effect transport.

The presence of injection barrier in the bulk hole transport provides an independent handle on the ambipolarity of the FET which is in the form of different dielectric interfaces. **Figure: 6a** shows the change in ambipolar transport with different dielectric materials in PFETs. The degree of ambipolarity (*D*) (= $\mu_{FET}^h/\mu_{FET}^e$) changes from ≈ 0.5 for BCB based devices to ≈ 0.92 for PVDF-HFP dielectric devices. This can be correlated to the presence of polar C-F groups at the interface in PVDF-HFP based PFETs which suitably alters the transport levels[41] to favor a balanced ambipolar transport (as shown in inset **Figure: 6a**). This interpretation is consistent with a |5 V| decrease in threshold voltage for hole transport when PVDF-HFP is used as the dielectric layer compared to low-*k* BCB based PFETs[42-43]. This element of control on the charge transport by modifying the dielectric layer is utilized to obtain high-performance and balanced ambipolar circuits.

### 3.7. Complementary Inverters

All-polymer complementary inverters were fabricated with this high performance polymer to demonstrate the usability in logic circuits. Inverters were characterized by the input-output ($V_{in}$ - $V_{out}$) transfer characteristics and gain curves. Idealistic (Z-type) voltage transfer curves were obtained with balanced charge transport in PVDF-HFP based dielectric layer. These ambipolar inverters with Au S-D electrode operated both in positive and negative input bias ($V_{dd}$). Voltage gain ($G = \frac{dV_{out}}{dV_{in}}$) magnitude of 40 was obtained with PVDF-HFP dielectric layer and 55 with low-*k* dielectric based inverter. However, the magnitude of *G* increases to 65 and 58 for *n*-FET load inverter and *p*-FET load inverter respectively upon using Al electrode in *n*-FET and Au S-D electrode in *p*-FETs. This enables better off state properties of the inverter, thus enhancing the gain. The operating frequency of these complementary logic circuits was obtained to be ~ 100 kHz by comparing the output voltage with the input pulse at various frequencies for devices with small channel length of 5 µm. Demonstration of these bipolar inverters with balanced transport high gain and fast switching response is a starting point in the design of 2DPP-TEG polymer based logic circuits.

## 4. Conclusion

N-type PFETs with $\mu_{FET}$ > 2 $cm^2V^{-1}s^{-1}$ was observed in 2DPP-TEG polymer system. A transition in *T* dependent processes from activated non-adiabatic transport to an extended adiabatic transport occurs in the high *T* regime where $\mu_{Hall} \sim \mu_{FET}$. The unifying requirement for obtaining polymeric systems with low degree of disorder is a combination of molecular and macroscopic parameter originating from co-planar structure which supports electron wave-function delocalization, interconnected aggregates along with an optimum interface and dielectric environment. Complementary inverters fabricated with this ambipolar polymer demonstrate high gain and fast switching response pointing to the advantages of ordered polymers for designing logic circuits. The results demonstrate a clear example of extended state transport in low-disordered polymeric materials which can form the basis to explore various phenomena expected from solution-processed high-mobility PFETs.

**Acknowledgements**

KSN and SPS acknowledge A.Sundaresan for measurements and discussions regarding Hall voltage, N.S.Vidyadhiraja for useful discussions, V.C.Kishore and C.Kulkarni for simulations. KSN acknowledges DAE, Government of India for partial funding. SPS acknowledges fellowship from CSIR, Government of India.

**Supplementary Section**

Supplementary section contains: evidence for 2D transport, structural analysis of the polymer, measurement of bias stress, water related trapping, analysis of dielectric-semiconductor interface width, dipolar contribution to $\mu_{FET}^{e}$ , GIXRD measurement, Annealing effect on transport, Details of Hall measurement, statistics of other devices.

**Figures:**

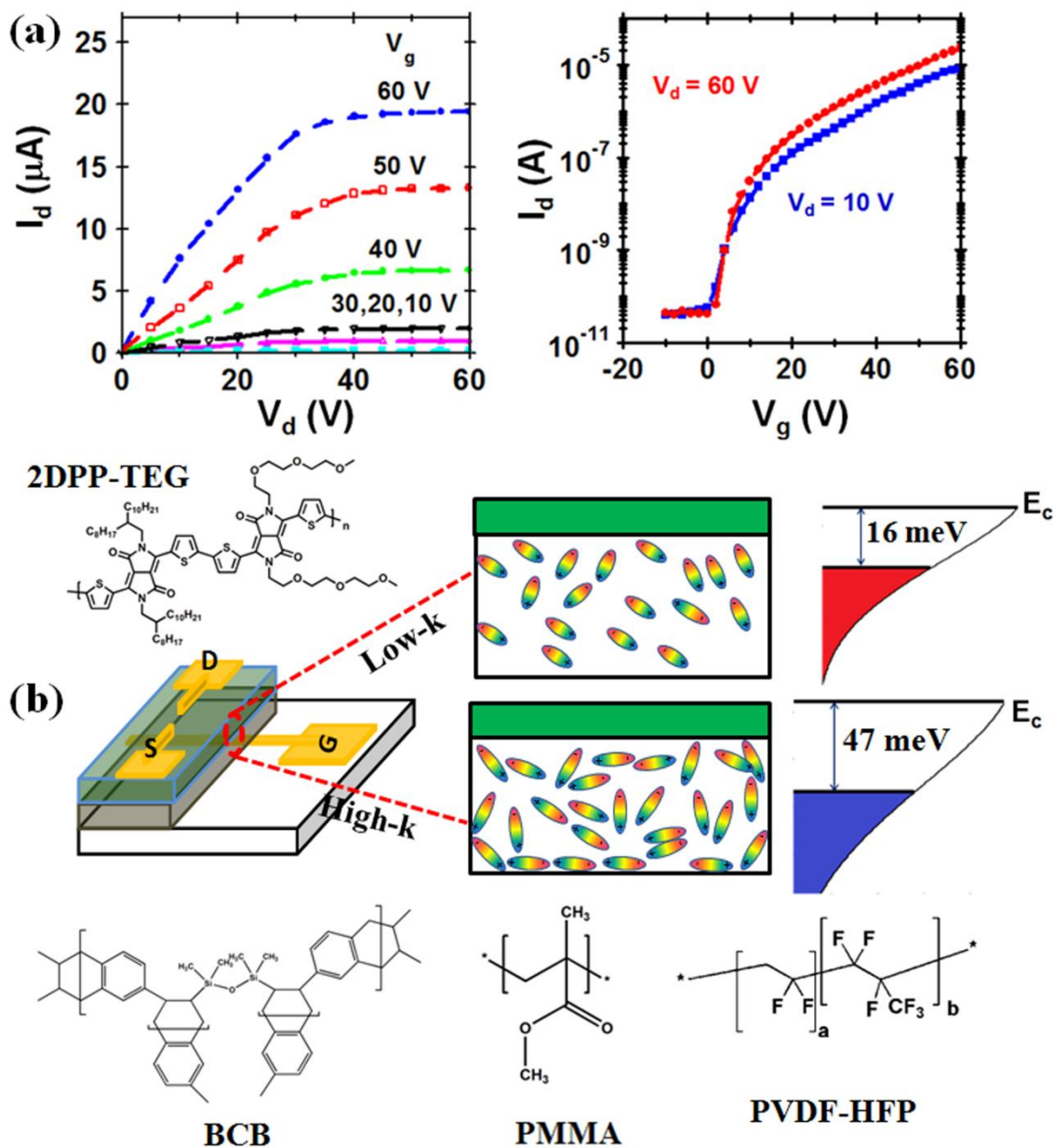

**Figure 1: PFET with 2DPP-TEG.** (a) Typical output and transconductance curve for a BCB based PFET with L = 60 μm and W = 1 mm; (b) schematic of the device and the DOS for semiconductor coated on low-k and high-k dielectrics, along with the chemical structure of the semiconducting polymer and dielectric polymers.

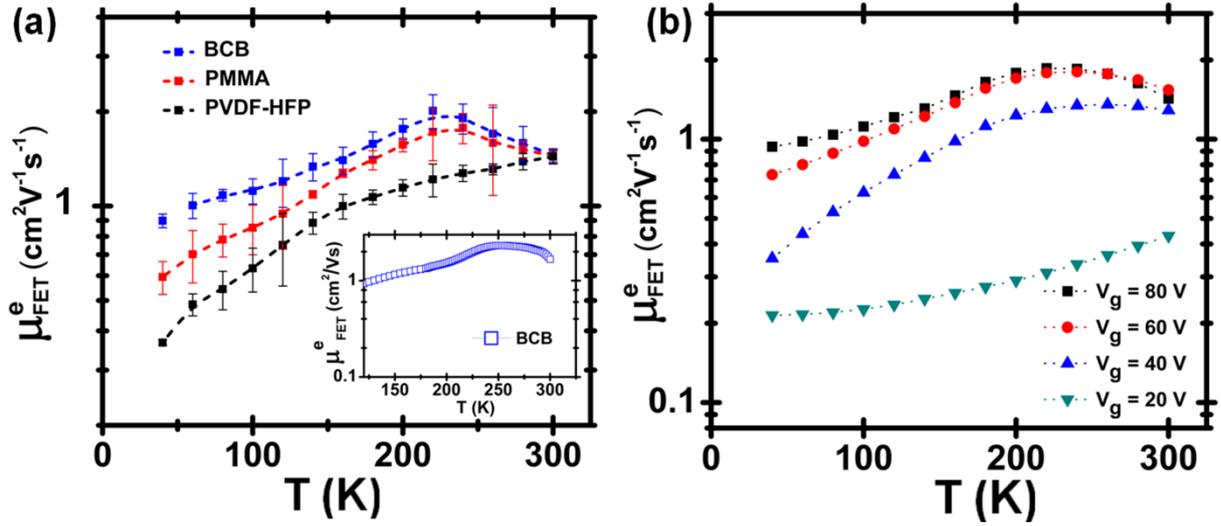

**Figure 2: T dependence of the PFET characteristics.** (a) $\mu_{FET}^{e}(T)$ of 2DPP-TEG FETs for different dielectrics; error bars indicate the mean deviation from a set of three devices; inset depicts $\mu_{FET}^{e}(T)$ with a higher resolution ($\Delta T \approx 2$ K) from a different batch of 2DPP-TEG sample fabricated on BCB dielectric; additional measurements of $\mu_{FET}^{e}(T)$ representing more number of devices is provided in **Supplementary Section S11**; (b) $V_g$ dependence of $\mu_{FET}^{e}(T)$ on a BCB based 2DPP-TEG FET.

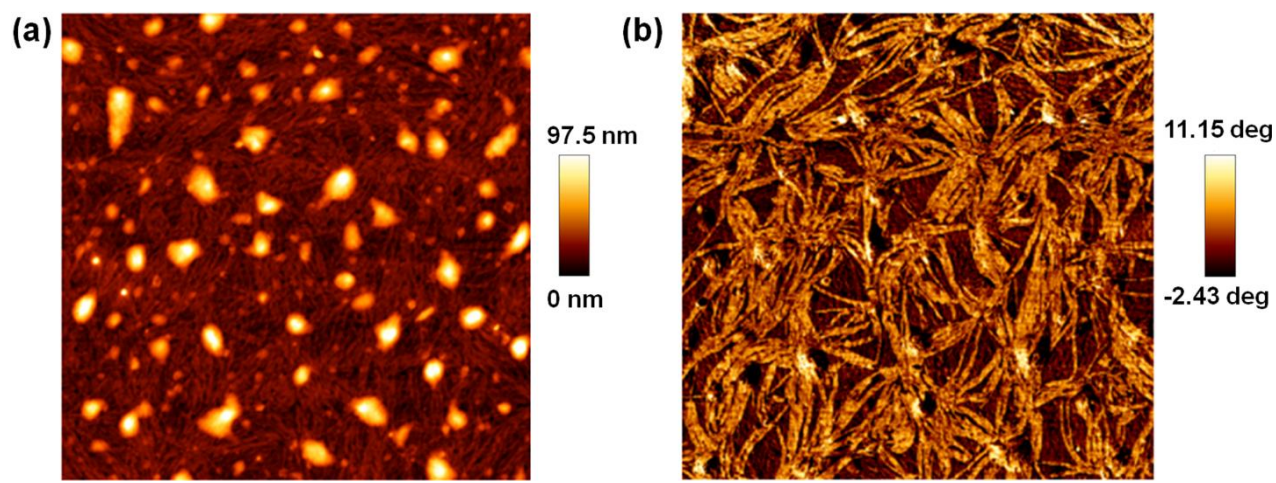

**Figure 3: Morphology of the polymer films.** (a) AFM topography and (b) phase-contrast image (10 μm × 10 μm) of the polymer film indicating self-assembled interconnected network of aggregates.

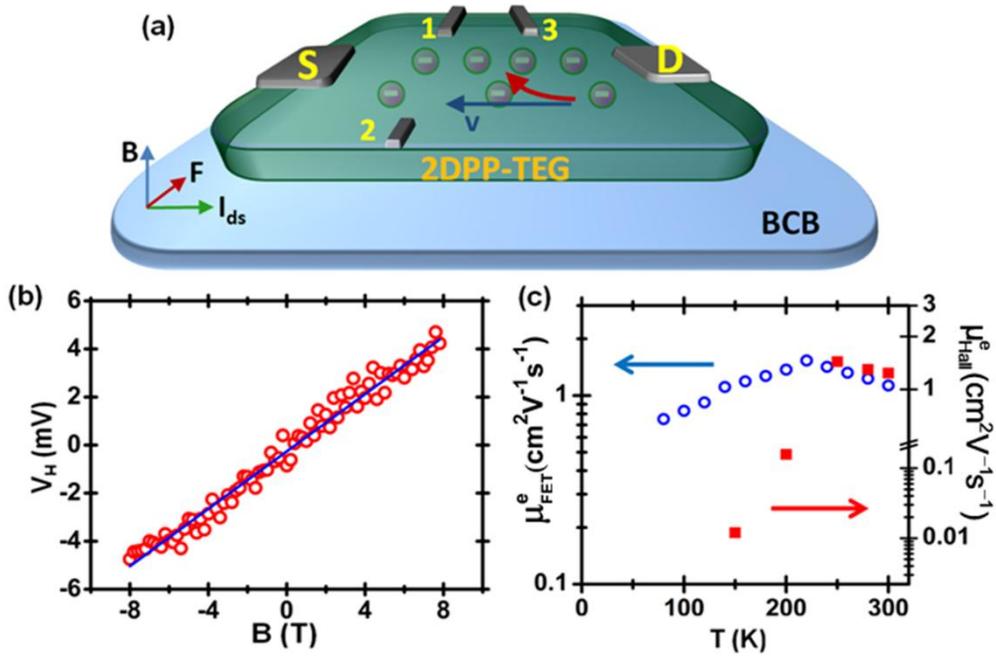

**Figure 4: Hall voltage measurements for 2DPP-TEG in FET geometry (L = 200 μm and W = 200 μm, W* = 40 μm and BCB dielectric).** (a) Schematic of the patterned Hall Voltage device geometry (patterned gate not shown), electrodes 1 and 2 represent the Hall probes and electrode 3 is for four-probe conductivity measurement; (b) Plot of $V_H$ with **B**; (c) comparative plot of $\mu^e_{FET}$ (blue markers) and $\mu^e_{Hall}$ (red markers) at different T obtained from another device.

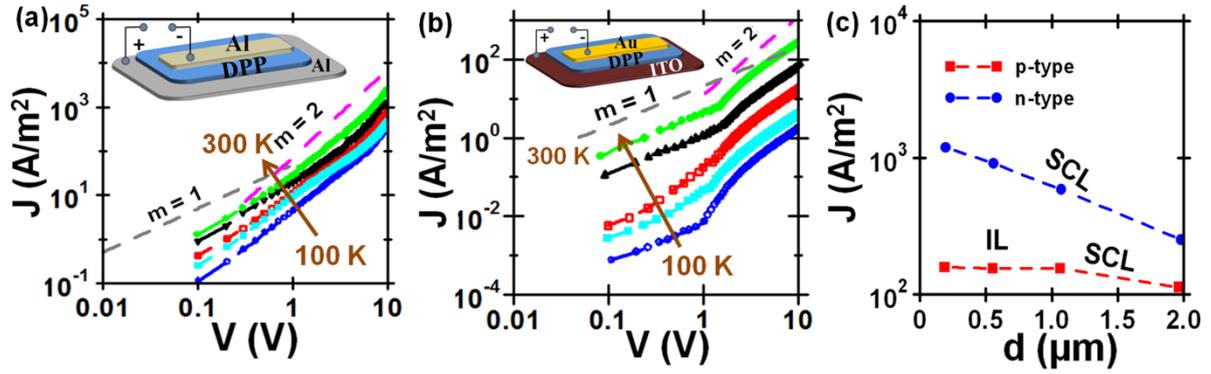

**Figure 5**: **Bulk characterization of 2DPP-TEG.** T dependent SCL transport measured for (a) electron-only device (d ~ 500 nm); (b) hole-only device (d ~ 2 μm). Inset shows the schematic of the devices. (c) Plot of J (**E** = 5 V/μm) with thickness of the active layer for both p-type and n-type transport showing injection limited (IL) and space charge limited (SCL) behavior.

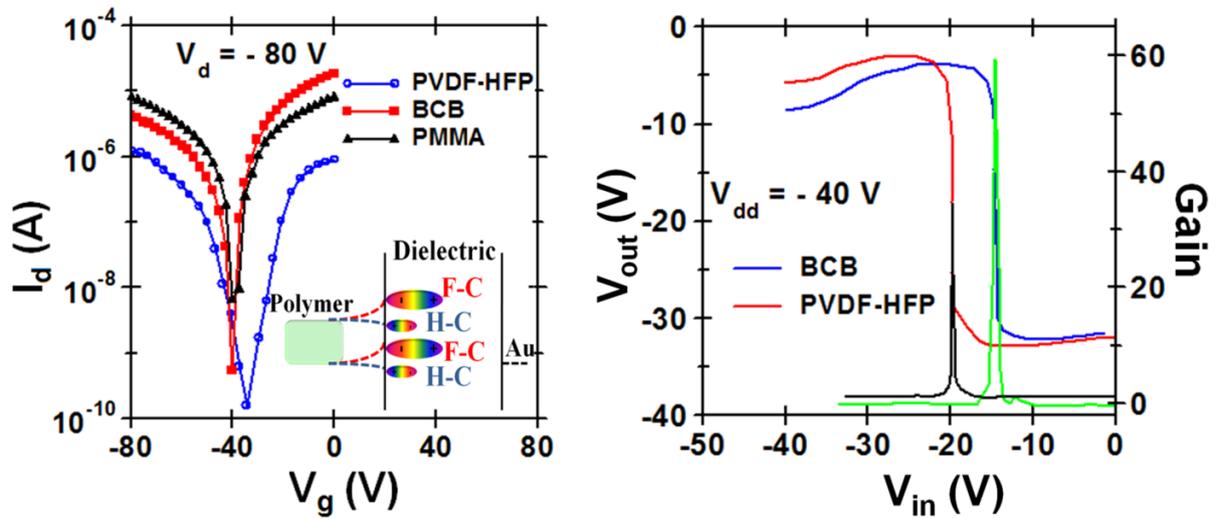

**Figure 6**: **Modification of ambipolar transport in 2DPP-TEG with different dielectric layers.** (a) Ambipolar transconductance curve for PFETs with L = 80 μm, W = 1 mm and different dielectric layers (inset shows the schematic of the energy levels of the MIS structure). (b) Typical voltage transfer and gain curves for inverters fabricated with 2DPP-TEG based ambipolar polymer using different dielectric layers.

| k | $\mu_{FET}$ (cm$^2$V$^{-1}$s$^{-1}$) at T$_{trans}$ | $\mu_{FET}$ (cm$^2$V$^{-1}$s$^{-1}$) at 300 K | E$_A$ (meV) | γ | D |
|---|---|---|---|---|---|
| 2.6 | 2.01 | 1.47 | 16 | 1.01 | 0.5 |
| 3.6 | 1.72 | 1.42 | 31 | 0.94 | 0.78 |
| 8 | - | 1.33 | 47 | - | 0.92 |

**Table 1:** Summary of the transport properties of 2DPP-TEG with different dielectrics fitted by a simple ME model.

# References:

[1] H. Sirringhaus, N. Tessler, and R. H. Friend, Science **280**, 1741 (1998).
[2] H. Klauk, Chem. Soc. Rev. **39**, 2643 (2010).
[3] X. Guo *et al.*, Nat Photon **7**, 825 (2013).
[4] I. I. Fishchuk, A. Kadashchuk, S. T. Hoffmann, S. Athanasopoulos, J. Genoe, H. Bässler, and A. Köhler, Physical Review B **88**, 125202 (2013).
[5] R. Noriega, J. Rivnay, K. Vandewal, F. P. V. Koch, N. Stingelin, P. Smith, M. F. Toney, and A. Salleo, Nat Mater **12**, 1038 (2013).
[6] R. A. Street, J. E. Northrup, and A. Salleo, Physical Review B **71**, 165202 (2005).
[7] J. Rivnay, R. Noriega, R. J. Kline, A. Salleo, and M. F. Toney, Physical Review B **84**, 045203 (2011).
[8] J. Zaumseil and H. Sirringhaus, Chem. Rev. **107**, 1296 (2007).
[9] T.-J. Ha, P. Sonar, and A. Dodabalapur, Phys. Chem. Chem. Phys. **15**, 9735 (2013).
[10] A. S. Dhoot, G. M. Wang, D. Moses, and A. J. Heeger, Phys. Rev. Lett. **96**, 246403 (2006).
[11] S. L. M. van Mensfoort and R. Coehoorn, Physical Review B **78**, 085207 (2008).
[12] S. Wang, M. Ha, M. Manno, C. Daniel Frisbie, and C. Leighton, Nat Commun **3**, 1210 (2012).
[13] M. J. Panzer and C. D. Frisbie, Adv. Funct. Mater. **16**, 1051 (2006).
[14] S. Stafstrom, Chem. Soc. Rev. **39**, 2484 (2010).
[15] C. Kanimozhi, N. Yaacobi-Gross, K. W. Chou, A. Amassian, T. D. Anthopoulos, and S. Patil, J. Am. Chem. Soc. **134**, 16532 (2012).
[16] J. Li *et al.*, Sci. Rep. **2** (2012).
[17] X. Zhang *et al.*, Nat Commun **4** (2013).
[18] T. Sakanoue and H. Sirringhaus, Nat Mater **9**, 736 (2010).
[19] T. Liu and A. Troisi, Adv. Funct. Mater., n/a (2013).
[20] J. Veres, S. D. Ogier, S. W. Leeming, D. C. Cupertino, and S. Mohialdin Khaffaf, Adv. Funct. Mater. **13**, 199 (2003).
[21] K. Asadi, A. J. Kronemeijer, T. Cramer, L. Jan Anton Koster, P. W. M. Blom, and D. M. de Leeuw, Nat Commun **4**, 1710 (2013).
[22] S. P. Senanayak, S. Guha, and K. S. Narayan, Phys. Rev. B **85**, 115311 (2012).
[23] J. H. Worne, J. E. Anthony, and D. Natelson, Appl. Phys. Lett. **96**, 053308 (2010).
[24] J. D. Yuen, R. Menon, N. E. Coates, E. B. Namdas, S. Cho, S. T. Hannahs, D. Moses, and A. J. Heeger, Nat Mater **8**, 572 (2009).
[25] U. Sondermann, A. Kutoglu, and H. Bassler, J. Phys. Chem. **89**, 1735 (1985).
[26] A. Salleo, T. W. Chen, A. R. Völkel, Y. Wu, P. Liu, B. S. Ong, and R. A. Street, Physical Review B **70**, 115311 (2004).
[27] H. L. Gomes, P. Stallinga, M. Cölle, D. M. de Leeuw, and F. Biscarini, Applied Physics Letters **88** (2006).
[28] L. Mattias Andersson, W. Osikowicz, F. L. E. Jakobsson, M. Berggren, L. Lindgren, M. R. Andersson, and O. Inganäs, Organic Electronics **9**, 569 (2008).
[29] S. Mehraeen, V. Coropceanu, and J.-L. Brédas, Physical Review B **87**, 195209 (2013).
[30] V. Podzorov, E. Menard, J. A. Rogers, and M. E. Gershenson, Phys. Rev. Lett. **95**, 226601 (2005).
[31] J.-F. Chang *et al.*, Phys. Rev. Lett. **107**, 066601 (2011).
[32] W. C. Germs, K. Guo, R. A. J. Janssen, and M. Kemerink, Phys. Rev. Lett. **109**, 016601 (2012).

[33] H. Yan, T. Schuettfort, A. J. Kronemeijer, C. R. McNeill, and H. W. Ade, Appl. Phys. Lett. **101**, 093308 (2012).
[34] N. A. Minder, S. Ono, Z. Chen, A. Facchetti, and A. F. Morpurgo, Adv. Mater. **24**, 503 (2012).
[35] Y. Xia, J. H. Cho, J. Lee, P. P. Ruden, and C. D. Frisbie, Adv. Mater. **21**, 2174 (2009).
[36] M. Caironi, M. Bird, D. Fazzi, Z. Chen, R. Di Pietro, C. Newman, A. Facchetti, and H. Sirringhaus, Adv. Funct. Mater. **21**, 3371 (2011).
[37] J. Takeya, K. Tsukagoshi, Y. Aoyagi, Takenobu, and I. Y, Jpn. J. Appl. Phys. **44**, L1393 (2005).
[38] E. K. Sichel, L. Greber, and K. Wang, Applied Physics Letters **52**, 1074 (1988).
[39] P. G. Le Comber, D. I. Jones, and W. E. Spear, Philosophical Magazine **35**, 1173 (1977).
[40] R. Steyrleuthner, M. Schubert, F. Jaiser, J. C. Blakesley, Z. Chen, A. Facchetti, and D. Neher, Adv. Mater. **22**, 2799 (2010).
[41] K.-J. Baeg, D. Khim, S.-W. Jung, M. Kang, I.-K. You, D.-Y. Kim, A. Facchetti, and Y.-Y. Noh, Adv. Mater. **24**, 5433 (2012).
[42] S. Kobayashi *et al.*, Nat Mater **3**, 317 (2004).
[43] S. A. DiBenedetto, A. Facchetti, M. A. Ratner, and T. J. Marks, Adv. Mater. **21**, 1407 (2009).

# Supporting information

# Room Temperature Band-like Transport and Hall Effect in a High Mobility Ambipolar Polymer

*Satyaprasad P.Senanayak[1], A.Z.Ashar[1], Catherine Kanimozhi[2], Satish Patil[2*] and K.S.Narayan[1*]*

Affiliation:

[1]Chemistry and Physics of Materials Unit

Jawaharlal Nehru Centre for Advanced Scientific Research

Bangalore 560064, India

[2] Solid State and Structural Chemistry Unit,

 Indian Institute of Science,

Bangalore 560012, India

*e-mail: narayan@jncasr.ac.in

satish@sscu.iisc.ernet.in

**Contents:**

## S1. 2-D transport and VRH treatment for transport

At a molecular level, we use the self-assembled interconnected aggregates as the basis for 2-D transport for PFETs fabricated with 2DPP-TEG polymer. These features are preserved for films of thickness 50 nm to 1 μm. Additionally, both top and bottom gated PFETs demonstrated similar $\mu_{FET}^e$ indicating identical morphology throughout the film.

In order to numerically prove the 2-D transport we perform the following analysis:

$$I_d^{2D} = A \frac{W}{L} d_{sc}^{1-\frac{T_0}{T}} \left(\frac{C}{e}\right)^{\frac{T_0}{T}} \frac{T}{T_0+T} (V_g - V_t)^{\frac{T_0}{T}+1}$$ for 2-D transport close to the interface and

$$I_d^{3D} = A \frac{W}{eL} d_{sc}^{1-\frac{T_0}{T}} \left(\frac{1}{2kT\epsilon\varepsilon_0}\right) C^{\frac{2T_0}{T}-1} \left(\frac{T}{2T_0-T}\right)\left(\frac{T}{2T_0}\right)(V_g - V_t)^{\frac{2T_0}{T}}$$ for 3-D transport where, $T_0$ is the characteristic width of the exponential DOS, $W$ is the channel width, $L$ is the channel length, $d_{sc}$ is the semiconductor thickness of constant carrier density, $C$ is the capacitance per unit area, $e$ is electronic charge and $A$ is related to carrier delocalization length and conductivity of the material (1).

In the saturation regime, both the 2D and 3D based model lead to direct power-laws:

$$I_d^{2D} \propto (V_g - V_t)^\alpha$$ where, $\alpha = \frac{T_0}{T} + 1$ for 2D transport and $\alpha = \frac{2T_0}{T}$ for 3D transport.

Analyzing the magnitude of $\alpha$ as a function of $T$ we determine the dimensionality of transport.

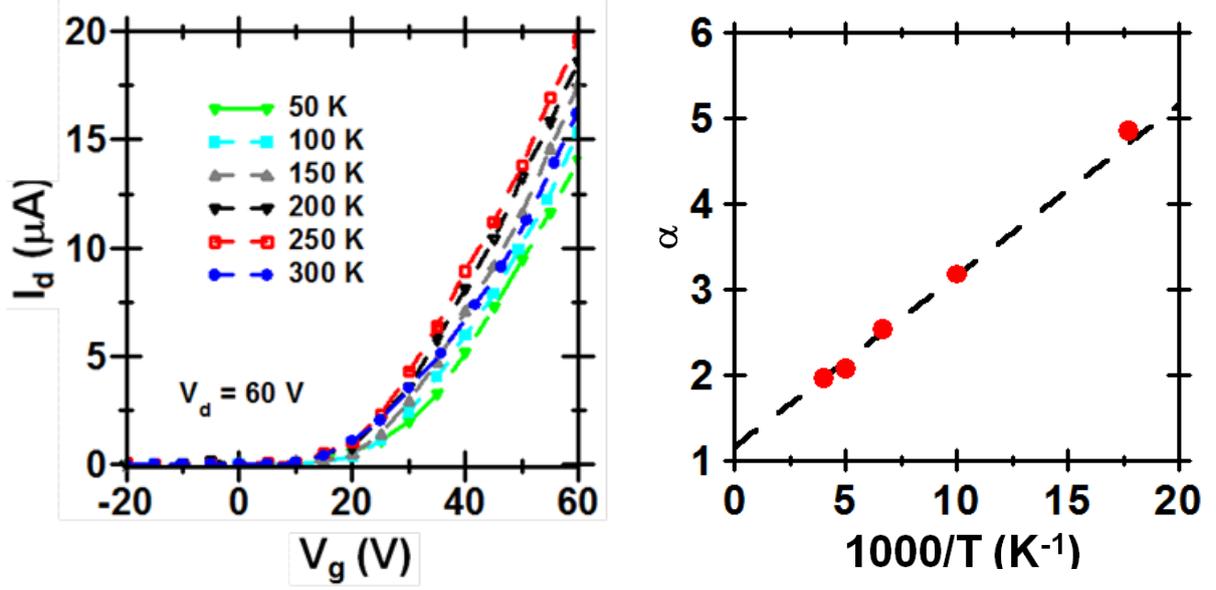

*Figure S1:* (a) Representative transfer characteristics of a 2DPP-TEG based FET ($L \approx 100$ μm, $W \approx 1$ mm) as a function of temperature. (b) Extracted values of $\alpha$ plotted versus $1/T$. The extrapolated linear fit yields the intercept with the vertical axis.

As evident from the plots of $\alpha$ v/s $1/T$ in the activated regime, the transport close to interface is 2D. Furthermore the observation of a finite Hall voltage supports the claim of transport phenomenon being not limited by the fibrillar 1-D structure and is governed by 2-D interconnected network (2).

We can model the gate induced order-disorder transition by variable range hopping in a 2-D network of localized states. As observed from **Figure: 2b** $\mu^{e}_{FET}(T)$ does not indicate a transport crossover at low $V_g$ ($\leq 20$ V) and at higher $V_g$ ($\geq 40$ V), $T_{trans}$ gets prominent indicating the crossover. This trend can possibly be related to the increase in the electron wave function delocalization with increase in gate induced charge density ($n_G$).

## S2. Structural Analysis of 2DPP-TEG polymer with T

In order to discount the effect of structural changes on the observed $\mu^e_{FET}$ trends dynamic scanning calorimetry (DSC) was performed on the molecule.

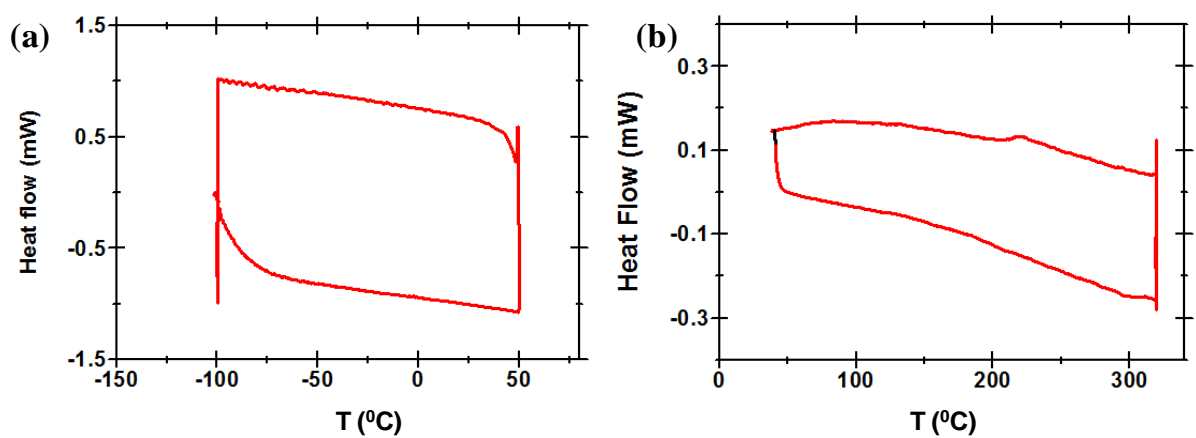

*Figure S2: DSC of the polymer 2DPP-TEG as taken from two different measuring instruments in the (a) temperature range (-100 °C - 50 °C) and (b) temperature range 40 °C - 320 °C.*

The measurement indicates no phase or structural transition in the range of -100 °C to 320 °C (173 K – 600 K) indicating that the origin of $T_{trans}$ (in the range of 200 K – 240 K) is purely because of transport mechanism variation. It should also be noted that the glass transition $T$ of the dielectric layers falls in the range of 100 °C for PMMA and 350 °C for BCB layer which is beyond the range of $\mu^e_{FET}(T)$ measurements.

FTIR transmission spectrum of 2DPP-TEG molecule was performed by embedding the molecule with anhydrous KBr forming a pellet. Temperature dependent FTIR transmission was obtained in the range of 160 K to 340 K in steps of 20 K under vacuum ($10^{-3}$ mbar) using BRUKER IFS 66v/S.

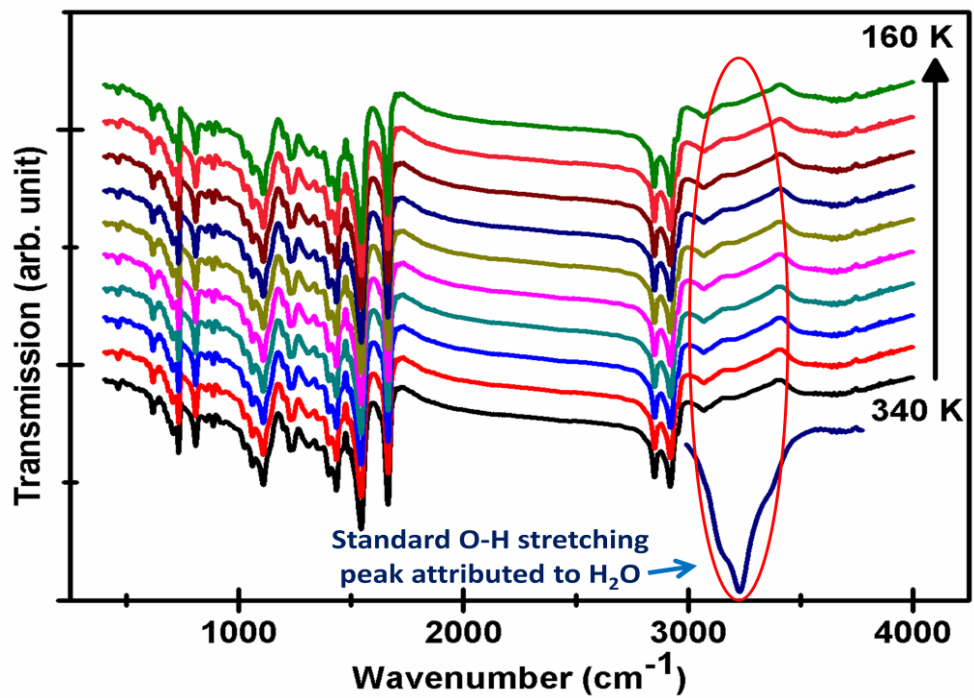

*Figure S3:* *FTIR measurement of the 2DPP-TEG in the T range of 160 K – 340 K. Also shown is a strong broad spectrum at ~ 3350 cm$^{-1}$ (attributed to H$_2$O) corresponding to O-H stretching.*

The FTIR spectra do not exhibit any structural transition in the complete *T* range. In general, the existence of absorbed/adsorbed water on the molecular surface is attributed to the simultaneous presence of ~ 3350 cm$^{-1}$ of O-H stretching and 1600 cm$^{-1}$ the scissoring frequency of water molecule. In the present case we do not observe a peak at 3350 cm$^{-1}$. Even though there is a transmission peak at 1600 cm$^{-1}$ (scissoring frequency of water) in each spectra, the absence of strong O-H stretching peak rules out the presence of water in the sample (3). Also the peak around 1600 cm$^{-1}$ is invariant with *T* which indicates that the peak is attributed to the sample itself.

## S3. Measurement of bias stress in 2DPP-TEG based PFETs

The negative coefficient in the $\mu^e_{FET}(T)$ can also originate due to bias stress in PFETs (4). Hence, PFET measurements were performed over a time scale of $10^4$ s. The device was continuously biased at $V_d$ = 60 V, $V_g$ = 60 V and transconductance sweep was performed at regular intervals.

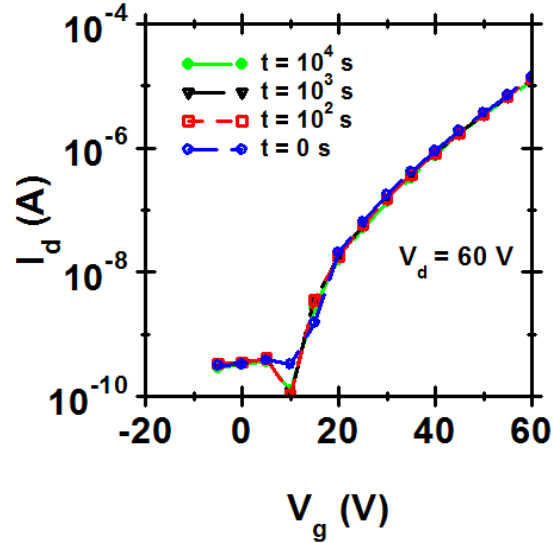

*Figure S4: Typical time variation of the transconductance measurement obtained for a PFET (L = 100 μm, W = 1mm) with BCB dielectric layer and 2DPP-TEG active layer.*

The transconductance curves demonstrate that the observed $\mu^e_{FET}(T)$ trends cannot be correlated to the bias stress in the device.

## S4. Water related trapping in PFETs

The observation of negative coefficient in the $\mu^e_{FET}(T)$ measurements can also originate from water related traps at the interface (5-7). Super-cooled $H_2O$ at the dielectric-semiconductor interface introduces traps which modifies the trends in the transport. In the present case this possible reasoning of water as a source for interpreting our observations is absolutely not possible. We lay out the following set of convincing arguments to prove this point:

(a) The procedure of measurement involved loading the samples on a probe station in a cryogenic vacuum assembly. The samples were heated to $T \sim 373$ K in high vacuum ($\sim 10^{-4}$ - $10^{-5}$ mbar) for a sufficiently long period (20-30 minutes) eliminating any possible sources of water in the sample and the chamber.

(b) Any form of temperature dependent measurements (conductivity, TGA, DSC, FTIR) does not show any storage of water molecules. To prove that the absorption of water molecule is negligible TGA (270 K – 700 K), DSC (200 K – 600 K) and FTIR (160 K- 240 K) measurements were performed (**reference** (8) **and Supporting Information S2**). TGA measurement does not demonstrate any weight loss for a $T \sim 600$ K and DSC measurements do not show any abnormal peaks relating to water absorption on the molecule. In addition FTIR also does not show water related absorption peaks in the complete $T$ range of 160 K – 340 K.

(c) Electrochemically, water and oxygen create trap due to the formation of oxidized polymer anions by the reaction (9):

$O_2 + 2H_2O + a\ Pol^-$ 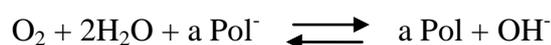 $a\ Pol + OH^-$

The reduction potential for half-cell involving water is $\approx -3.6$ eV. Electron trap levels in polymer due to water is generally observed at energy levels of -3.6 eV (10). Hence molecules with LUMO level deeper than -3.6 eV are stable to water related trapping (9) which is indeed the case with 2DPP-TEG (8). The over-potential at the LUMO of 2DPP-TEG makes the process energetically unfavorable. Additionally, degradation of organic molecules can also arise due to percolation of water molecules inside the polymer chain. This can be avoided by molecular design which enables tighter packing and prevents the percolation of water molecule (9, 11-12). From the molecular point of view, the amphiphillic substitution in 2DPP-TEG provides stronger aggregates ($\pi$- $\pi$ stacking established by GIXRD experiments (13)); hence the role of water molecule affecting the charge transport is minimal.

(d) The hydrophillicity of the polymer surface was probed and the water contact angle was measured to be ($110^0 \pm 7^0$) proving the hydrophobic nature of the polymer surface.

(e) A sizable number of devices (> 100) were measured. The characteristics of the devices does not depend on the history of the devices (Storage-exposure-cooling without heating), clearly pointing out the absence of effects from residual moisture.

(f) It should also be noted that the reports indicating super-cooled water effects revealed changes in slope for $I_{ds}(T)$ or $\mu^e_{FET}(T)$ on a smaller $T$ range (14-15) beyond which the activated behavior is revived. However, the $\mu^e_{FET}(T)$ trend in 2DPP-TEG shows variation in slope which continues till a $T$ range of 300 K much above the super-cooled phase transition temperature.

(g) Magnitude of the $T_{trans}$ is dependent on the gate voltage $V_g$. If trapping due to super-cooled water is responsible for the negative coefficient of mobility, then the features should be more pronounced at low $V_g$, as opposed to what is observed in the present case (**fig 2b** of the manuscript).

Detailed measurements were also performed to point out that the intrinsic nature of transport is controlled by the structure of semiconducting polymer and the interfacial energetics. Each of these aspects was verified by tweaking the molecular structure and the dielectric layer respectively. For example: Range of measurements was performed on 2DPP-2Dod (the polymer analog without the order promoting ethylene glycol substitution) based FETs. Typical mobility magnitude was obtained to be $10^{-3}$ cm$^2$V$^{-1}$s$^{-1}$ and the $\mu^e_{FET}(T)$ behavior does not exhibit the crossover. The $E_a$ obtained from $\mu^e_{FET}(T)$ of 2DPP-2Dod was significantly high and is ≈ 100-150 meV. The other trend observed in the $\mu^e_{FET}(T)$ of 2DPP-TEG is that the transition temperature ($T_{trans}$) can be tuned with the dielectric layer on which the polymer layer is coated (**Figure: 2a**). It was observed that with increase in polarity of the dielectric layer $T_{trans}$ increases to higher value which results in the suppressing the band-like features (as shown in **Figure: 2a** of the manuscript). This trend seen in FETs fabricated with PVDF-HFP dielectric layer can be directly attributed to the interface traps introduced with polar dielectrics. It should be noted that the possible source of water from the dielectric component was also completely eliminated by combination of thermal and vacuum treatments. Contact angle measurements were also performed on dielectric surfaces which points to a hydrophobic surface. These set of observations provide adequate evidence that $\mu_{FET}^e(T)$ trends originate from the inherent transport mechanisms and not from artifacts due to external factors.

Additionally, in order to ascertain that super-cooled water does not affect the $\mu_{FET}^e(T)$ measurements in our set up we also obtained low-$T$ transport measurements on a range of other polymers: P3HT and N2200.

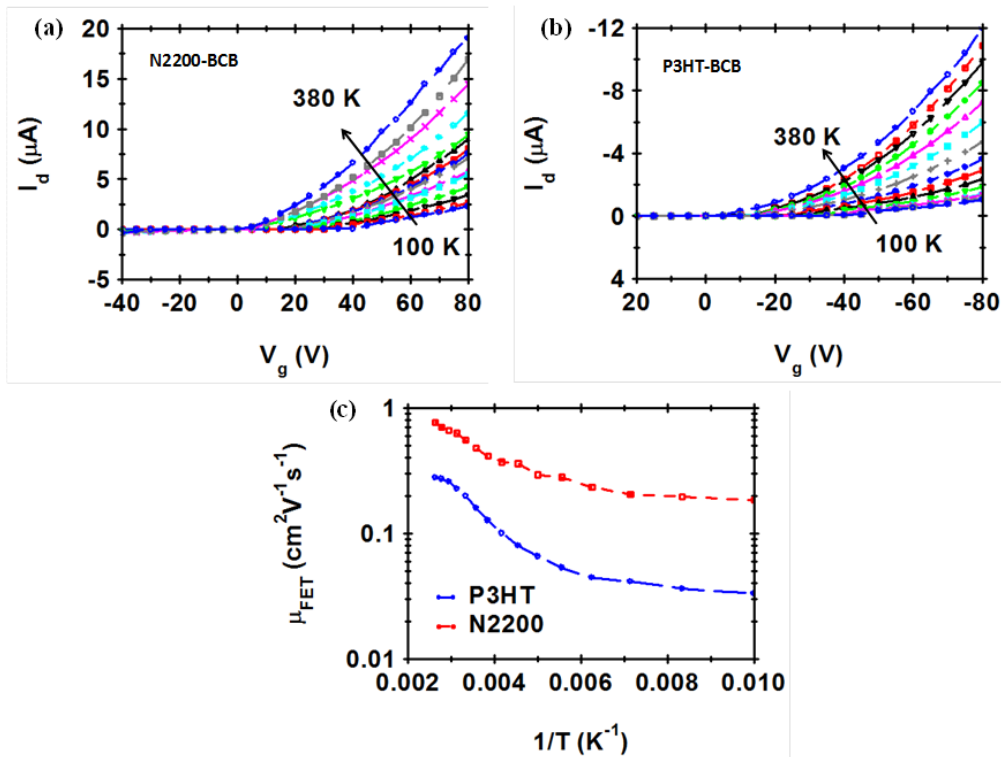

*Figure S5:* *Typical transconductance plots for PFETs fabricated with BCB dielectric layer and (a) N2200 (n-type) or (b) P3HT (p-type) as the semiconducting layer. Typical dimensions of channel were L = 100 μm and W = 1 mm; (c) $\mu_{FET}^e(T)$ variation for both FETs.*

As evident from the figures no abnormal discontinuities in the $\mu^e_{FET}(T)$ behavior was observed for any of the molecules, nevertheless the activation energies observed were comparable to the best reported values for the respective polymers. These observations clearly show that the problems pertaining to super-cooled water do not exist in our measurement set-up.

The above mentioned trends do lead us to conclude that the possibility of super-cooled water as the origin and source for the observed features is not justified.

## S5. Analysis of dielectric-semiconductor interface width

In order to ascertain that the change in the transport behavior is reflective of the dipolar disorder at the interface and not because of changes like interface de-mixing, modification of interface due to solvent or change in interfacial width, AFM morphology of the dielectric layer were measured before and after solvent treatment. The dielectric layer was exposed to similar chemical and annealing treatment as used during device fabrication and the morphology was monitored. For the imaging, JPK Nanowizard 3 with APPNANO ACTA Aluminium coated Si-cantilevers (force constant 40 Nm$^{-1}$ and resonance frequency $f \sim 375$ kHz) was used in non-contact mode.

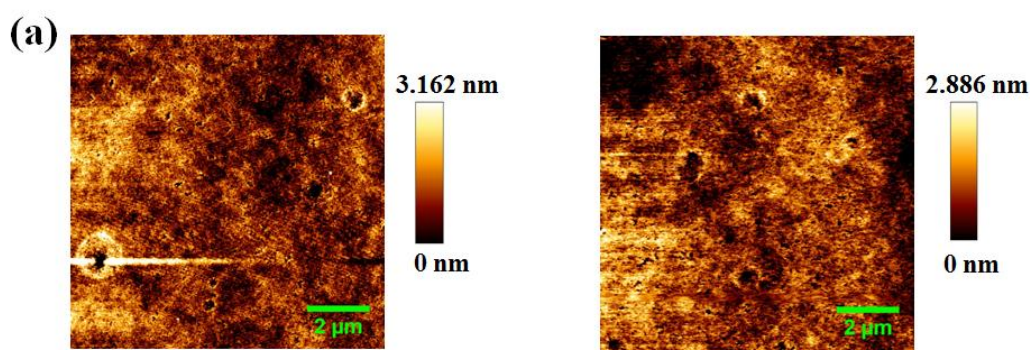

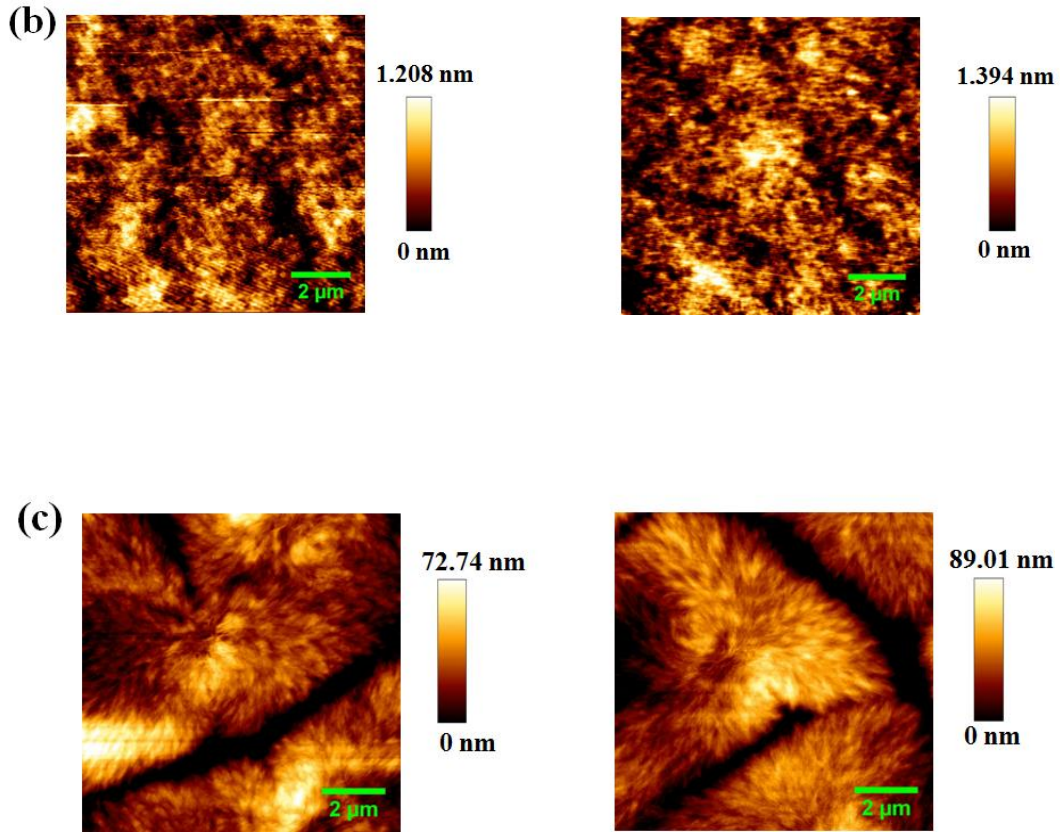

*Figure S6:* AFM morphology of the dielectric layers: (a) BCB (b) PMMA (c) PVDF-HFP before (left) and after (right) solvent treatment.

**Table T1**: Roughness parameters obtained from AFM of the dielectric films.

| Dielectric | Treatment | Average roughness (nm) | r.m.s roughness (nm) |
|---|---|---|---|
| BCB | Before | 0.52 | 0.71 |
| | After | 0.51 | 0.65 |
| PMMA | Before | 0.25 | 0.31 |
| | After | 0.23 | 0.28 |
| PVDF-HFP | Before | 12.05 | 16.53 |
| | After | 15.48 | 20.23 |

As can be seen from the table the roughness parameters do not vary significantly with solvent treatment indicating that the trends in $\mu_{FET}^e(k)$ originates from dipolar disorder and has minimal contribution from physical changes in the interface.

## S6. Dipolar contribution to mobility

To quantitatively, estimate the contribution of polarization induced disorder on the mobility we use a simplistic approach based on electrostatics interaction of the potential due to the dielectric layer and the organic molecule. An important ingredient of transport model is the DOS of the materials. In the present case exponential DOS is assumed for modeling the transport. In the presence of random dipolar potential due to dielectric layer the shape of the DOS is modified. The potential $V$ can be assumed to have a Gaussian distribution with probability, $P(V) \propto exp(-e^2V^2/2E_p^2)$; where, $P(V)$ is the probability distribution of the potential due to the dielectric dipoles, $E_p$ is the energetic disorder generated due to the random dipoles of the dielectric layer. Due to the presence of this random dipolar potential the exponential DOS close to the conduction band gets modified and we assume it to be converted to Gaussian DOS(16). Hence, the modified DOS is then given by: $\rho(E) = \frac{N_G}{\delta_G^2} \exp\left[-\frac{(E-E_c)^2}{2\delta_G^2}\right]$ where, $N_G$ is the total number of states, $\delta_G$ is the width of the DOS and $E_c$ is the conduction band. This broadening corresponds to the increased depth of the tail states which results in decreased mobility of the polymer. So we estimate the dielectric induced disorder by the simplistic formula: $\delta_G = \sqrt{E_{sc}^2 + E_p^2}$ where $E_{sc}$ is the inherent disorder in the semiconductor. We obtain $E_{sc}$ by assuming that the transport in BCB is weakly affected by the disorder. Using this we estimate the broadening of the DOS in 2DPP-TEG due to dipoles of PMMA and PVDF-HFP ≈ 26 meV and 44 meV respectively.

## S7. GIXRD of 2DPP-TEG

2DPP-TEG thin film was evaluated using grazing-incidence X-ray diffraction (GIXRD), and the scattering pattern is shown in **Figure: S7**. The polymer adopts a common lamellar structure and edge-on packing with a coherence length of 9.1 nm. The π-π stacking (010) peak has an associated *d* spacing of 0.36 nm and a coherence length of 3.4 nm.

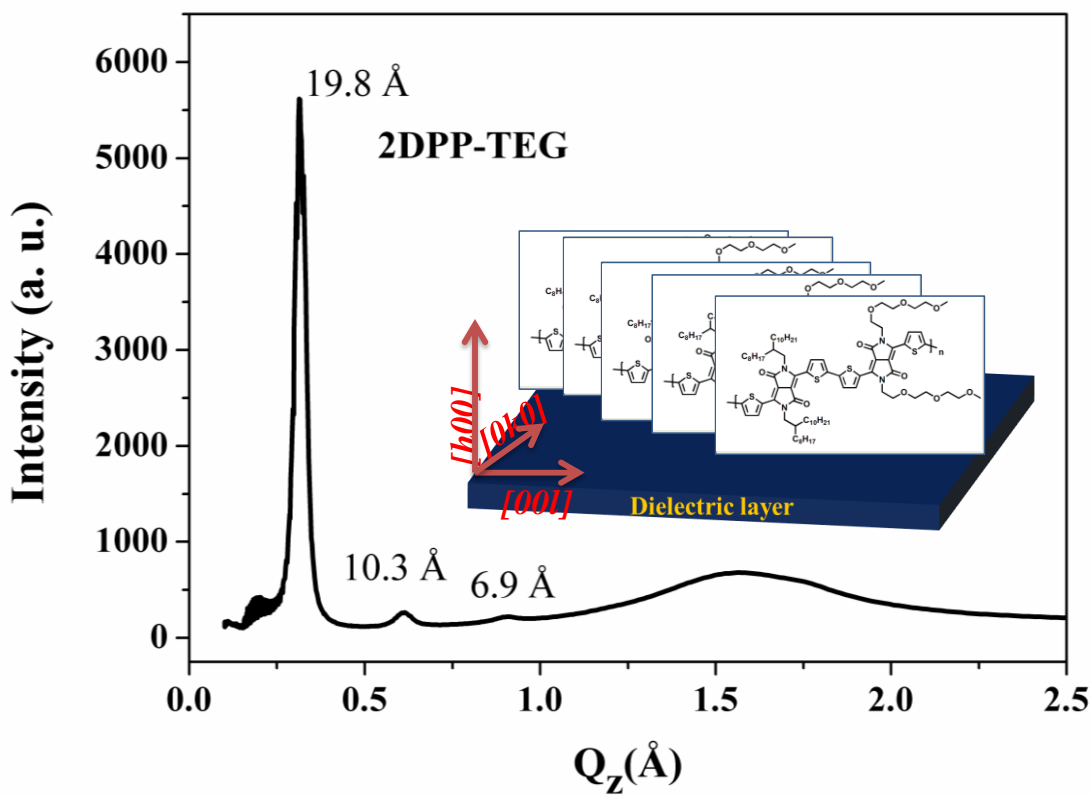

*Figure S7: GIXRD pattern for 2DPP-TEG thin films, exhibiting prominent peaks corresponding to (100) and π-π stacking. Inset shows the schematic for edge-on stacking of 2DPP-TEG on the dielectric layer.*

GIXRD spectra indicate edge-on stacking of the semiconducting core on the dielectric layer. This result in the semiconducting conjugated core electrostatically isolated from the dielectric induced disorder due to the alkyl chain and TEG substitution

## S8. Role of Annealing on electron transport

Annealing of polymer thin films in a PFET modifies the crystallinity of the polymer and also minimizes the trap density at the interface. To understand the effect of annealing on the electron and hole transport FETs were fabricated with different annealing *T* ranging from 100 °C to 200 °C and transport was measured. For electron transport, Al S-D electrodes were used and for hole transport Au S-D electrodes were used. It was observed that with increase in annealing *T*, the $\mu_{FET}^{e}$ increases from 1.1 cm$^2$V$^{-1}$s$^{-1}$ to 2.5 cm$^2$V$^{-1}$s$^{-1}$ whereas $\mu_{FET}^{h}$ remains fairly constant in the range of 0.2 – 0.5 cm$^2$V$^{-1}$s$^{-1}$. To understand this variation AFM measurement were performed for films annealed at different *T*.

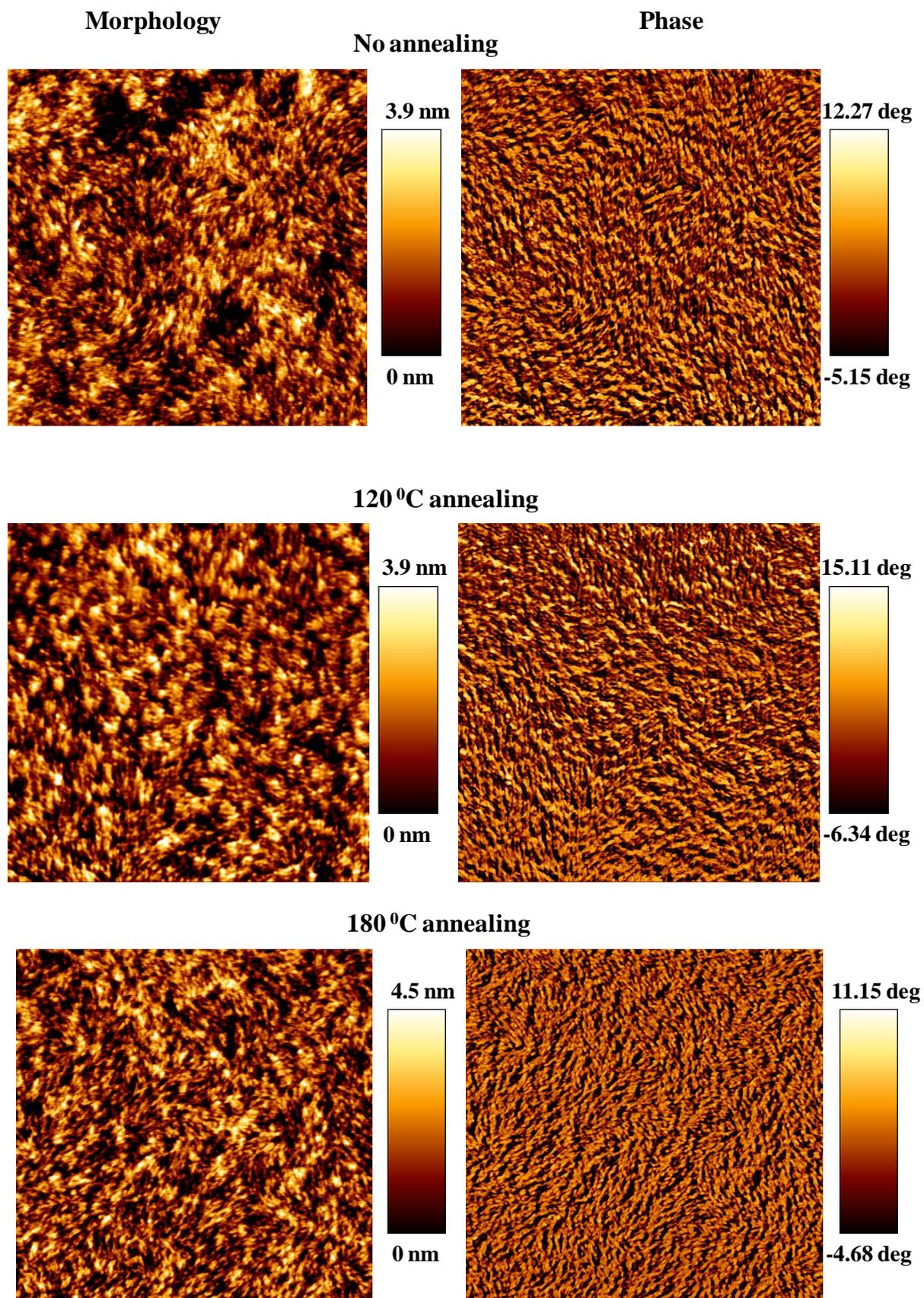

*Figure S8:* *AFM images (3 µm × 3 µm) of the semiconducting film annealed at different T.*

As seen from the AFM images annealing of the films does not vary the grain size and similar interconnected network is maintained throughout. Additionally, any modification in grain or morphology should be reflected both in electron and hole transport. Since there is minimal change in hole transport with annealing, we believe that the modification of transport with annealing can be mainly attributed to the decrease in electron trap density at the interface. These measurements demonstrate the critical role of interface traps in the transport.

## S9. Characterization of controlled polymers

To understand the importance of $\pi$-$\pi$ stacking and amphiphillic design on the observation of band like behavior controlled polymer 2DPP-2Dod (DPP with no ethylene glycol substitution and two Do-decyl substitutions) was synthesized. The $\mu_{FET}^e$ magnitude was obtained to be $10^{-3}$ cm$^2$V$^{-1}$s$^{-1}$ for the controlled polymer. Additionally, the molecule also shows decreased crystallinity and no self-assembled inter-connected microstructures as evident from XRD and SEM images.

For the XRD image thin films of the polymers were prepared under similar condition as used for the device fabrication. SEM images for the films (drop-casted from a solution of 2mg/ml concentration) were obtained using Zeiss Ultra 555 at an operating voltage of 5kV.

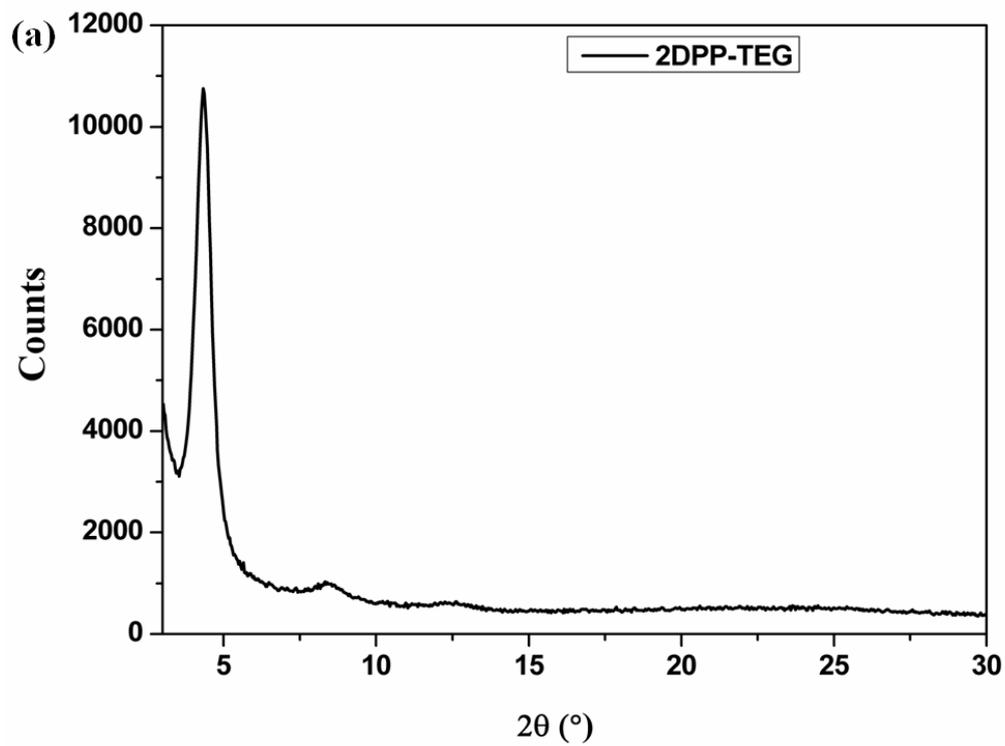

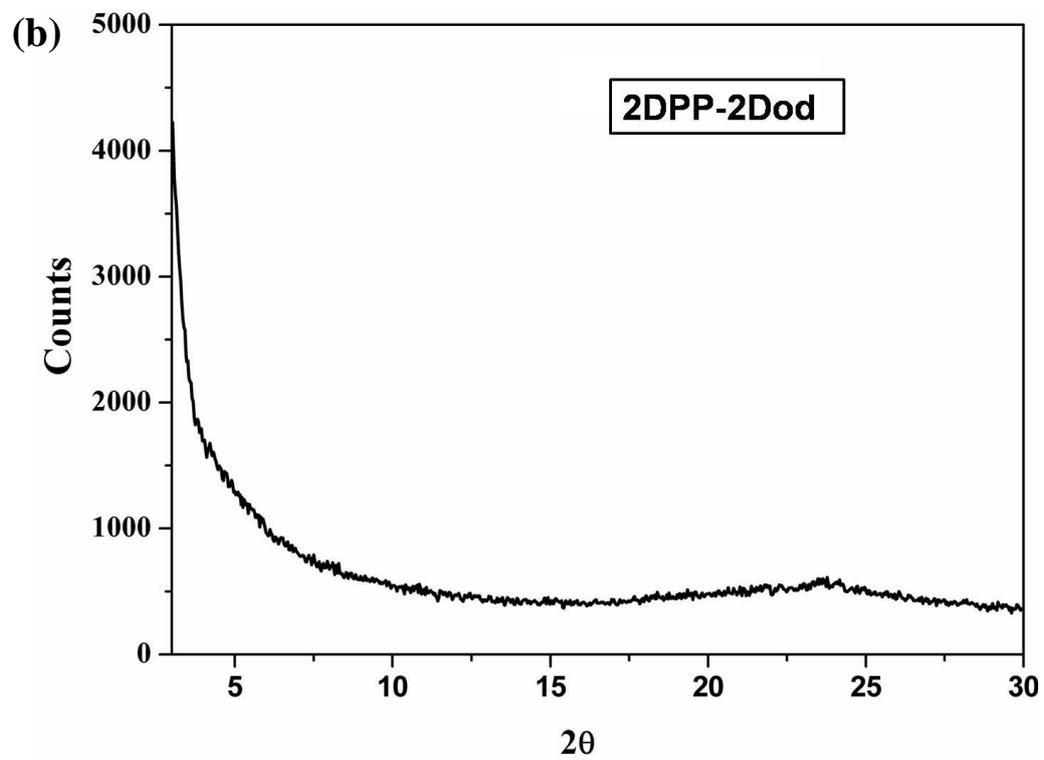

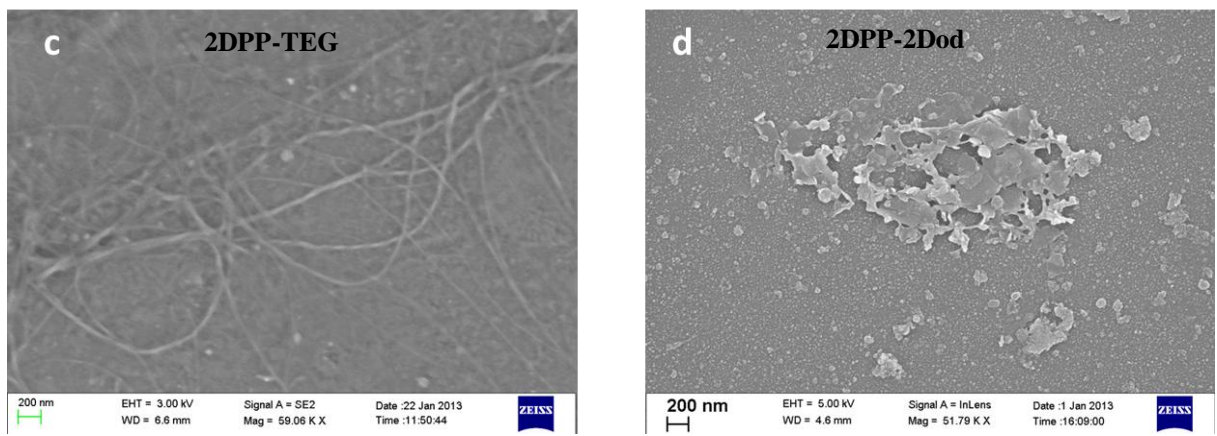

*Figure S9:* XRD images of (a) 2DPP-TEG and (b) 2DPP-2Dod thin films. SEM images obtained from very dilute solutions of (c) 2DPP-TEG and (d) 2DPP-2Dod thin films.

FET measurements on these controlled polymer indicate that for observation of high $\mu_{FET}^{e}$ and band-like behavior both enhanced $\pi$ electron overlap and tight packing due to amphiphillic substitution is essential (13).

## S10. Hall measurement

Hall Voltage measurements were performed on PFET device architecture by ramping the *B* between + 8 T to − 8 T for multiple cycles in steps of 0.1 T or 0.2 T at a slow rate of 0.2 T/min using a home-built probe assembly in a physical property measurement system from Quantum Design Inc. Keithley 2001 Multimeter (input impedance ∼ 10 TΩ) was used to measure $V_H$ (after correction of the drift) while the PFETs were biased ($V_d$ = 10 V and $V_g$ = 60 V) using Keithley 2400 source meters. In the PFETs selected to measure the Hall voltage it was ensured that $\mu_{FET}^{e}$ values were in the range of 1- 3 cm$^2$V$^{-1}$s$^{-1}$ with a discernible negative coefficient of temperature dependence of the mobility.

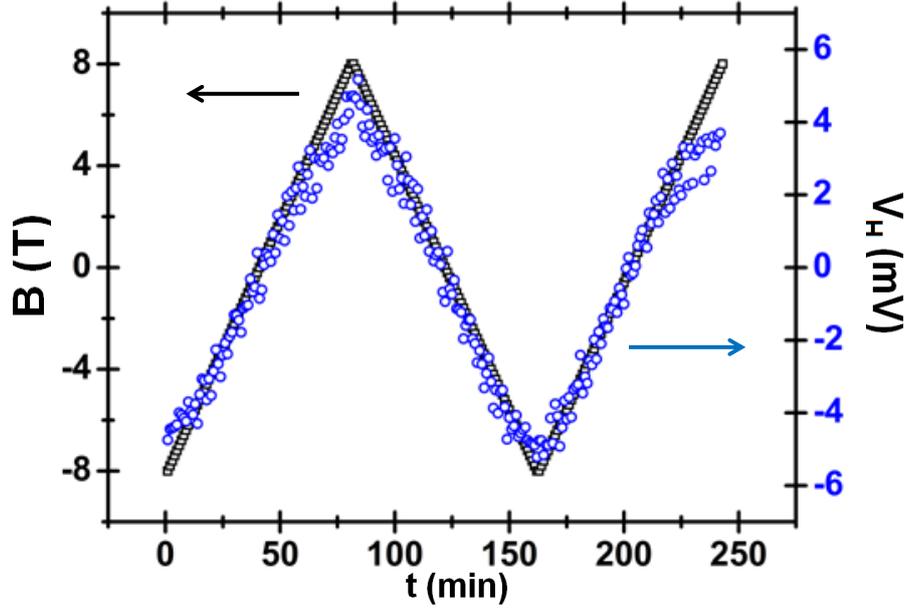

**Figure S10:** *Plot of $V_H$ with **B** for 2DPP-TEG in a PFET (L = 200 μm and W = 200 μm, W\* = 40 μm and BCB dielectric) under bias conditions $V_d$ = 10 V, $V_g$ = 60 V measured over multiple cycles of **B** variation.*

As evident from the figure fairly consistent **B** dependence of the $V_H$ is obtained for multiple cycles of **B** variation. In general, $V_H$ varies linearly with $V_{ds}$ due to increase in the $I_{ds}$. To obtain more insight from the Hall measurement, devices were fabricated with varied dimensions of channel and Hall electrodes. Typical device dimensions utilized for the measurement were maintained in the range of L = 200 - 400 μm, W = 200 - 500 μm and the distance between Hall electrodes were varied in the range of 200 μm - 1 mm with width of 40 μm - 80 μm. In general, it was ensured that the *L* is 5 to 10 times of the width of Hall probes (*W\**). This ensures proper detection of Hall voltage at various position of the channel. In addition, *L\** was varied to have Hall probes within the channel as well as outside the channel. Hall probes inside the channel can result in additional conducting pathways (17), hence results in little higher $V_H$ magnitude (variation of 20 % after geometrical correction factor).

The mean deviation in the estimation of $n_H$ and $\mu_{Hall}$ were obtained from different geometries (six devices) and was estimated to be $\pm 0.2 \times 10^{12}$ cm$^{-2}$ and $\pm 0.5$ cm$^2$V$^{-1}$s$^{-1}$ respectively.

## S11. Statistics of $\mu_{FET}^e$(T) trends on a set of devices:

$\mu_{FET}^e(T)$ were obtained on a range of devices fabricated from 2DPP-TEG polymer. The polymer was synthesized in different batches. Measurements were made on each of the batches. Typical $\mu_{FET}^e$ values were in the range of 1- 3 cm$^2$V$^{-1}$s$^{-1}$. A continuous transition in $\mu_{FET}^e(T)$ is observed in devices with $\mu_{FET}^e > 1$ cm$^2$V$^{-1}$s$^{-1}$ as shown in the accompanying figures below.

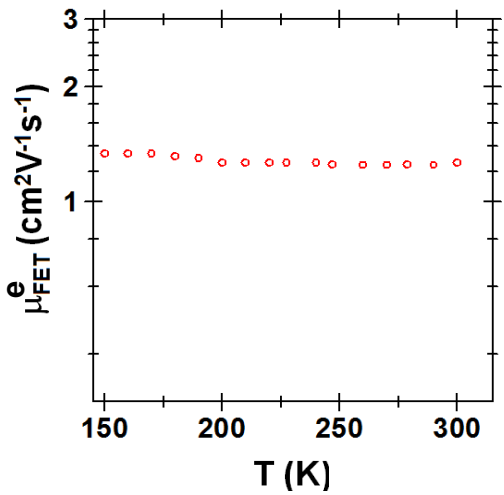
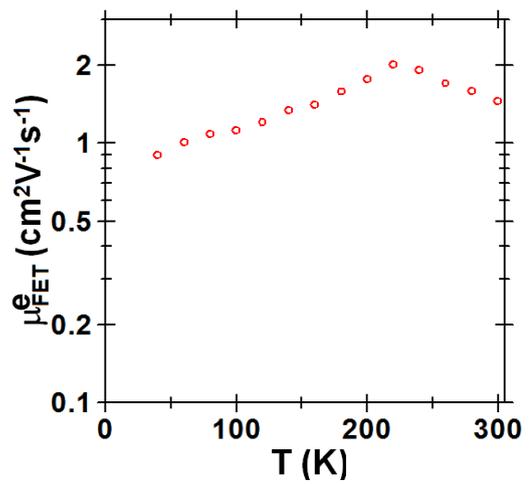
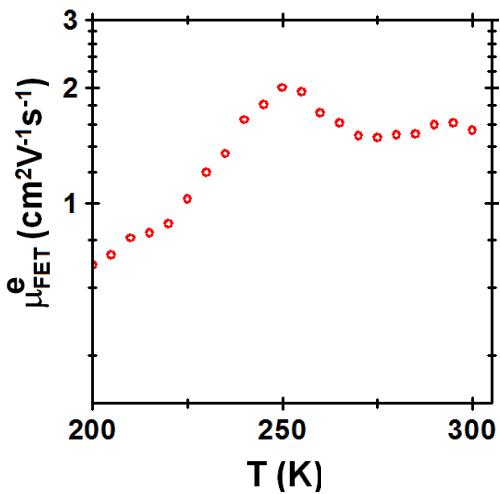
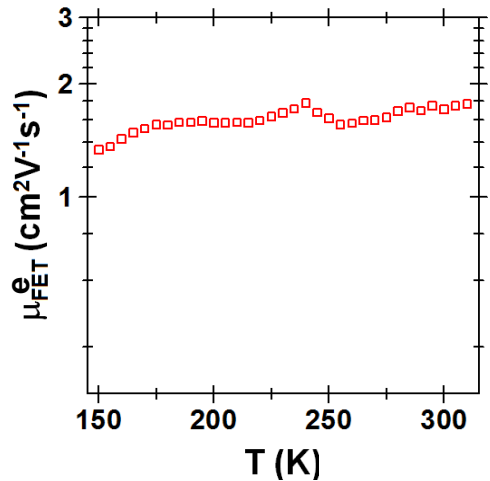

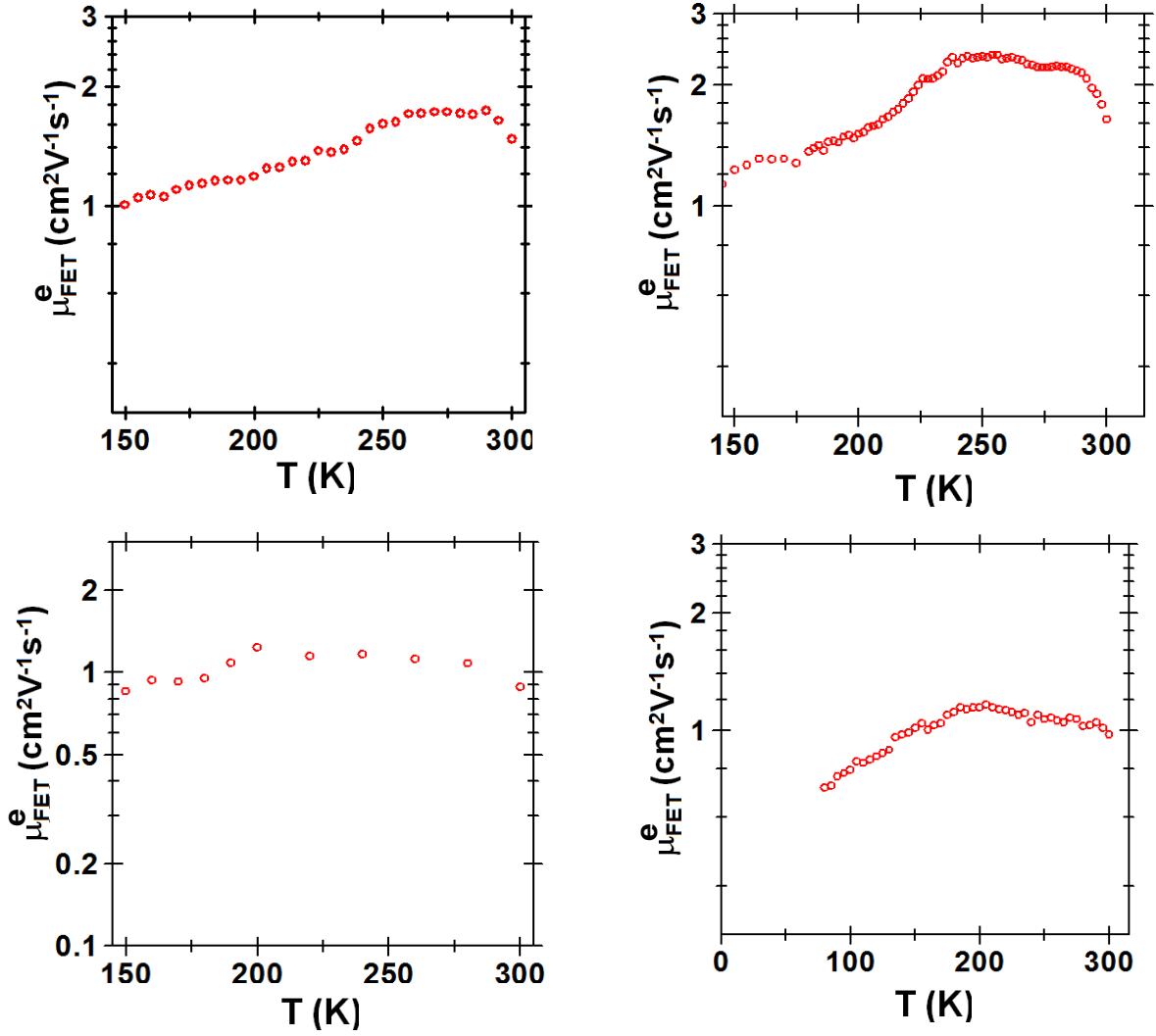

*Figure S11:* Representative $\mu^e_{FET}(T)$ measurements for a range of PFETs fabricated with different batches of 2DPP-TEG and BCB dielectric layers. Typical L values were in the range of (60 – 100) μm and W ≈ 1 mm. Magnitude of room temperature $\mu^e_{FET}$ and low T regime $E_a$ were comparable for a wide selection of devices. (Total number of devices tested > 160).

# References:

1. Kronemeijer AJ, *et al.* (2014) Two-Dimensional Carrier Distribution in Top-Gate Polymer Field-Effect Transistors: Correlation between Width of Density of Localized States and Urbach Energy. *Adv. Mater.* 26(5):728-733.
2. Yuen JD, *et al.* (2009) Nonlinear transport in semiconducting polymers at high carrier densities. *Nat Mater* 8(7):572-575.
3. Braun Dietrich HC, Matthias Rehahn, Helmut Ritter, Brigitte Voit. (2012) Polymer Synthesis: Theory and Practice: Fundamentals, Methods, Experiments (5th edition). .
4. Salleo A, *et al.* (2004) Intrinsic hole mobility and trapping in a regioregular poly(thiophene). *Physical Review B* 70(11):115311.
5. Street RA, Northrup JE, & Salleo A (2005) Transport in polycrystalline polymer thin-film transistors. *Physical Review B* 71(16):165202.
6. Gomes HL, Stallinga P, Cölle M, de Leeuw DM, & Biscarini F (2006) Electrical instabilities in organic semiconductors caused by trapped supercooled water. *Applied Physics Letters* 88(8):-.
7. Mattias Andersson L, *et al.* (2008) Intrinsic and extrinsic influences on the temperature dependence of mobility in conjugated polymers. *Organic Electronics* 9(5):569-574.
8. Kanimozhi C, *et al.* (2012) Diketopyrrolopyrrole–Diketopyrrolopyrrole-Based Conjugated Copolymer for High-Mobility Organic Field-Effect Transistors. *J. Am. Chem. Soc.* 134(40):16532-16535.
9. Di Pietro R, Fazzi D, Kehoe TB, & Sirringhaus H (2012) Spectroscopic Investigation of Oxygen- and Water-Induced Electron Trapping and Charge Transport Instabilities in n-type Polymer Semiconductors. *J. Amer. Chem. Soc.* 134(36):14877-14889.
10. Nicolai HT, *et al.* (2012) Unification of trap-limited electron transport in semiconducting polymers. *Nat Mater* 11(10):882-887.
11. Jones BA, Facchetti A, Wasielewski MR, & Marks TJ (2007) Tuning Orbital Energetics in Arylene Diimide Semiconductors. Materials Design for Ambient Stability of n-Type Charge Transport. *J. Amer. Chem. Soc.* 129(49):15259-15278.
12. Mei J & Bao Z (2013) Side Chain Engineering in Solution-Processable Conjugated Polymers. *Chem. Mat.*
13. Kanimozhi C, *et al.* (2013) Role of Alkyl Chain in Rational Design of N-Type Diketopyrrolopyrrole-based Conjugated Polymers: What have we learned? in *Manuscript under preparation*.
14. Gomes HL, Stallinga P, Cölle M, de Leeuw DM, & Biscarini F (2006) Electrical instabilities in organic semiconductors caused by trapped supercooled water. *Appl. Phys. Lett.* 88(8):-.
15. Kemerink M, *et al.* (2009) Temperature- and density-dependent channel potentials in high-mobility organic field-effect transistors. *Phys. Rev. B* 80(11):115325.
16. Mehraeen S, Coropceanu V, & Brédas J-L (2013) Role of band states and trap states in the electrical properties of organic semiconductors: Hopping versus mobility edge model. *Phys. Rev. B* 87(19):195209.
17. David JM & Buehler MG (1977) A numerical analysis of various cross sheet resistor test structures. *Solid-State Electronics* 20(6):539-543.